\documentclass[lettersize,journal]{IEEEtran}
\usepackage{amsmath,amsfonts}
\usepackage{algorithmic}
\usepackage{algorithm}
\usepackage{array}
\usepackage{textcomp}
\usepackage{stfloats}
\usepackage{url}
\usepackage{verbatim}
\usepackage{graphicx}
\usepackage{cite}
\usepackage{booktabs} 
\usepackage{amsthm, amssymb}
\usepackage[font=normalsize, textfont=sf, justification=centering]{caption}
\usepackage{subfigure}
\usepackage{fancyhdr}
\newcommand{\tabincell}[2]{\begin{tabular}{@{}#1@{}}#2\end{tabular}}  
\begin{document}
	\title{Resource Management for IRS-assisted WP-MEC Networks with Practical Phase Shift Model }
   \author{Nana Li, Wanming Hao,~\IEEEmembership{Member,~IEEE,}
     	Fuhui Zhou,~\IEEEmembership{Senior Member, IEEE},
		Zheng Chu,~\IEEEmembership{Member,~IEEE}, 
		Shouyi Yang, and Pei Xiao,~\IEEEmembership{Senior Member, IEEE}
		\thanks{N. Li, W. Hao and S. Yang are with the School of Electrical and Information Engineering, Zhengzhou University, Zhengzhou 450001, China. (e-mail: nnli@gs.zzu.edu.cn, iewmhao@zzu.edu.cn, iesyyang@zzu.edu.cn)}
				\thanks{F. Zhou is with the College of Electronic and Information Engineering, Nanjing University of Aeronautics and Astronautics, Nanjing, 210000, China. (e-mail: zhoufuhui@ieee.org)}
		\thanks{ Z. Chu and P. Xiao are with the Institute for Communication Systems, University of Surrey, Guildford GU2 7XH, UK. (e-mail: andrew.chuzheng7@gmail.com, p.xiao@surrey.ac.uk)} }

%

\IEEEpubid{\begin{minipage}{\textwidth}\ \\ \\ \\ \\ \\ \centering 
	\copyright 2023 IEEE. Personal use of this material is permitted. However, permission to use this material for any other purposes must \\ be obtained from the IEEE by sending a request to pubs-permissions@ieee.org.
\end{minipage}}
\maketitle
    
  \begin{abstract}
    Wireless powered mobile edge computing (WP-MEC) has been recognized as a promising solution to enhance the computational capability and sustainable energy supply for low-power wireless devices (WDs). However, when the communication links between the hybrid access point (HAP) and WDs are hostile, the energy transfer efficiency and task offloading rate are compromised. To tackle this problem, we propose to employ multiple intelligent reflecting surfaces (IRSs) to WP-MEC networks. Based on the practical IRS phase shift model, we formulate a total computation rate maximization problem by jointly optimizing downlink/uplink IRSs passive beamforming, downlink energy beamforming and uplink multi-user detection (MUD) vector at HAPs, task offloading power and local computing frequency of WDs, and the time slot allocation. Specifically, we first derive the optimal time allocation for downlink wireless energy transmission (WET) to IRSs and the corresponding energy beamforming. Next, with fixed time allocation for the downlink WET to WDs, the original optimization problem can be divided into two independent subproblems. For the WD charging subproblem, the optimal IRSs passive beamforming is derived by utilizing the successive convex approximation (SCA) method and the penalty-based optimization technique, and for the offloading computing subproblem, we propose a joint optimization framework based on the fractional programming (FP) method. Finally, simulation results validate that our proposed optimization method based on the practical phase shift model can achieve a higher total computation rate compared to the baseline schemes. 
   \end{abstract}
		
	\begin{IEEEkeywords}
			Wireless powered mobile edge computing (WP-MEC), intelligent reflecting surface (IRS), phase shift model, resource management.
	\end{IEEEkeywords}
	
	\section{Introduction}
	\subsection{Background}
	\IEEEPARstart{I}n future wireless networks, a large number of internet of things (IoTs) wireless devices (WDs) are envisioned to be interconnected \cite{ref1, ref2}. However, due to the constraints of manufacturing cost and device size, WDs are typically equipped with life-limited batteries and energy-saving low-performance processors, which imposes a great challenge to perform the computing-intensive tasks with a low latency \cite{ref3, ref4}. Recently, wireless powered mobile edge computing (WP-MEC), which integrates the MEC and wireless energy transmission (WET) \cite{ref5, ref6, ref7, ref8}, has emerged as a promising technique to solve this issue. In WP-MEC systems, a hybrid access point (HAP) equipped with an edge server can simultaneously  provide wireless charging and computing services for WDs. However, the performance of the WP-MEC systems may not be guaranteed when the channels between the WDs and HAPs are blocked by buildings or other obstacles.
    
    To tackle the above issue, intelligent reflecting surface (IRS) can be deployed to provide an additional transmission link for WDs. In general, an IRS comprises of a controller circuit and a large number of low-cost reflection elements \cite{ref9}. Regulated by the IRS controller, the reflection angle of incident signal can be intelligently adjusted by optimizing the phase shift of each element, improving the energy transfer efficiency and task offloading rate. However, the designs of IRS elements are most based on the ideal phase shift model where the signals can be fully reflected by each element regardless of its phase shift, which is an unrealistic assumption. Generally, the reflection amplitude is dependent on the reflection phase shift \cite{ref10}, and an optimal balance should be found between the reflection amplitude and phase shifter to maximize the reflection efficiency.  Therefore, in this paper, we consider an IRS-assisted WP-MEC system based on the practical phase shift model.

   \subsection{Related Works}
   Recently, there have been several works investigating the resource optimization in IRS-assisted MEC networks and IRS-assisted WP-MEC networks.

  \subsubsection{IRS-assisted MEC Networks}
  The potential of MEC cannot be fully unleashed when the communication links between HAPs and WDs are poor. Therefore, some works proposed to bring IRS to MEC systems for improving the task offloading rate. For example, Bai \emph{et al}. investigated the latency-minimization problem subject to practical constraints imposed on the edge computing capability \cite{ref11}. To exploit the spare computing resources of IoT devices, the authors of \cite{ref12} aimed to minimize the overall latency and proposed a device-to-device (D2D) cooperative computing strategy. In addition to computing latency, the authors of \cite{ref13} focused on maximizing the sum computational bits, a key indicator to evaluate the computing capability for the IRS-assisted MEC networks. Furthermore, the authors of \cite{ref14} considered how to minimize the energy consumption of devices with the help of IRS. Additionally, MEC also provided a platform for artificial intelligence (AI) applications thanks to the rich computation resources to train machine learning (ML) models, resulting in minimizing learning errors for all devices \cite{ref15}. In order to reduce the computational complexity of conventional optimization methods, the authors of \cite{ref16} proposed a learning-based algorithm to achieve online task offloading and resource management in the IRS-assisted MEC networks.

  \subsubsection{IRS-assisted WP-MEC Networks} In \cite{ref11, ref12, ref13, ref14, ref15, ref16}, IRSs are only used to improve the data transmission rate between HAPs and IoT devices. To realize the simultaneously energy and data transmission, Chen \emph{et al}. proposed to maximize the computation rate subject to the constraint of wireless energy received by WDs in a multi-user scenario, where the time division multiple access (TDMA) technology is applied to avoid the multi-user interference \cite{ref17}. Following this, Chu \emph{et al}. aimed to maximize the network utility by jointly optimizing the HAP's transmit power and IRS passive beamforming based on the TDMA \cite{ref18}. Different from the above work \cite{ref17, ref18}, considering the single-antenna HAP, the authors of \cite{ref19, ref20, ref21} considered the multiple antenna HAP and investigated the joint HAP active and IRS passive beamforming design in the IRS-assisted WP-MEC system. Specifically, the authors of \cite{ref19} formulated a delay minimization problem by jointly optimizing the IRS phase shift, edge computing resource, HAP MUD matrix, energy/data transmission time, and task offloading coefficient of each WD. Similarly, Li \emph{et al}. considered the latency minimization problem and proposed a multiple access scheme for multi-user task offloading \cite{ref20}. Meanwhile, the authors of \cite{ref21} aimed to maximize the sum computation rate in the proposed multiple IRSs-assisted WP-MEC works and the optimal time scheduling and computing models were obtained via the one-dimensional search method and greedy algorithm.  
  
  Based on the above analysis, although there are several works jointly investigating MEC and WET in the IRS-assisted networks, most of them are based on the ideal IRS reflection model and without taking into account the IRS energy consumption, which is not practical. Therefore, in this paper, we consider the practical IRS reflection model, where the reflection amplitude is dependent on the reflection phase shift. Meanwhile, multiple IRSs and HAPs are deployed to jointly serve WDs, and IRSs can obtain the energy via the multiple HAPs energy beamforming. In this case, we aim to maximize the total computation rate of all WDs by jointly optimizing  downlink/uplink IRS passive beamforming, HAP downlink energy beamforming and uplink MUD vector, task offloading power and local CPU-cycle frequency of WDs, and time allocation for WET and data computing. To the best of our knowledge, it is the first time to investigate such model. Our work is significantly different from the works of Xiaowei Pang mainly from three aspects. First, most of the author's work focus on addressing the problems encountered in UAV communication, and the IRSs are exploited to mitigate interference \cite{ref22}, facilitate secure transmission \cite{ref23}, or target sensing \cite{ref24}, while our work investigates the IRS-assisted WP-MEC networks.  Second, we adopt a practical phase shift model rather than an ideal model to capture the reflection characteristics of IRS. Third, due to the different problem formulations, we propose a new algorithm to cater to the new formulated problem.
   
  \subsection{Contributions}  
  Against this background, we consider a multiple IRS-assisted WP-MEC network with a practical IRS phase shift model, and investigate the resource scheduling strategy to coordinate the wireless charging for IRSs and WDs, and task offloading for WDs. The main contributions are detailed as follows.
	\begin{itemize}
		\item[•] We consider a multiple IRS-assisted WP-MEC network with a practical IRS phase shift model. Under this framework, IRSs first receive wireless energy  from HAPs, and then utilize the harvested energy to assist both the downlink energy transfer to WDs and uplink task offloading. Meanwhile, multiple HAPs and IRSs  cooperatively serve WDs. 
		  
		\item[•] We formulate a total computation rate maximization problem by jointly optimizing IRSs downlink/uplink passive beamforming, HAP downlink energy beamforming and uplink MUD vector, task offloading power and local CPU-cycle frequency of WDs, and time slot allocation for WET and data computing under the energy casuality constraints of IRSs and WDs. We first derive the optimal time allocation for downlink energy transfer to IRSs and the corresponding energy beamforming, Next, with fixed time allocation for WDs charging, the original optimization problem can be divided into two independent subproblems. 

		\item[•] The first subproblem aims to optimize the settings of WET to WDs, in which the corresponding downlink energy beamforming and passive beamforming design is a feasibility-check problem. Next, we reformulate this problem to maximize the energy harvested by WDs via introducing a set of auxiliary variables. Then, an alternative optimization (AO) is proposed to obtain the HAPs energy beamforming and IRSs passive beamforming. The second subproblem aims to optimize the settings of data computing. We propose a joint optimization framework to find a suboptimal solution of the computing setting efficiently. Specifically, we iteratively optimize the MUD vector, IRSs passive beamforming, and task offloading power with the aid of the Lagrangian dual reformulation (LDR) and \emph{Multidi mensional Complex Quadratic Transform} (MCQT) technology.
		
		\item[•] Simulation results demonstrate that compared to the scheme without IRS, the IRS can significantly improve the performance of the WP-MEC network. Furthermore, the performance loss caused by the imperfect hardware of IRS is more pronounced. Therefore, it is necessary to consider the practical IRS phase shift model for the passive beamforming design in the IRS-assisted communication systems.
	\end{itemize} 
	
	The remainder of this paper is organized as follows. We describe the system model and formulate a total computation rate maximization problem in Section II. The solution of the proposed problem is provided in Section III. Section IV presents the simulation results. Finally, Section V concludes this paper.
	
	\begin{table}[thbp]
		\centering
		\footnotesize
		\renewcommand{\arraystretch}{1.5}
		\setlength{\tabcolsep}{3pt}
		\caption{Notation List.}
		\scalebox{.91}{
			\begin{tabular}{ll}
				\toprule
				\textbf{Parameters} & \textbf{Definition}  \\ \midrule
				 $ K $  & Number of WD  \\
				 $ B $    & Number of HAP  \\
			     $M$ &  Number of HAP's antennas \\
				 $ I $        &  Number of IRS  \\ 
			     $ N $  & Number of IRS reflection element \\
				 $ P_{max} $        & Maximum transmit power of HAP \\
				 $ \mathbf{h}_{b,k}^{d} $  & Direct channel from the $ b $-th HAP to the $ k $-th WD \\
				 $ \mathbf{G}_{b,i} $  & Reflected channel from the $ b $-th HAP to the $ i $-th IRS \\
				 $ \mathbf{h}_{i,k}^{r} $ & Reflected channel from the $ i $-th IRS to the $ k $-th WD \\
				 $ \mathbf{h}_{k}^{d} $  & Direct channel from HAPs to the $ k $-th WD \\
				 $ \mathbf{G}_{i} $  & Reflected channel from HAPs to the $ i $-th IRS \\
				 $ \mathbf{h}_{k}^{r} $ & Reflected channel from IRSs to the $ k $-th WD \\
				 $ \mathbf{h}_{k} $  & Overall channel from HAPs to the $ k $-th WD \\
				 $ \beta_{i,n} $ & Reflecting amplitude at the $ n $-th reflecting element of the $ i $-th IRS \\
				 $ \theta_{i,n} $ & Phase shift at the $ n $-th reflecting element of the $ i $-th IRS  \\
				 $ \mathbf{\Theta}_{i} $ & Phase shift matrix at the $ i $-th IRS  \\
				 $ \mathbf{\Theta} $ & Overall Phase shift matrix at all IRSs \\
                 $ \mathbf{v} $  & Passive beamforming and $\mathbf{v}=\mathbf{\Theta}\mathbf{1}_{IN}$ \\
				 $ \beta_{min} $ & Minimum reflecting amplitude in the practical phase shift model \\
				 $ \mathbf{s} $ & Transmit signal of WDs \\
				 $ \mathbf{u}_{k} $ & Uplink multi-user detection vection of the $ k $-th WD \\
				 $ \mathbf{W} $ & Energy transmit covariance matrix at the first sub-slot $ \tau_{1} $ \\
				 $ \mathbf{Q} $ & Energy transmit covariance matrix at the first sub-slot $ \tau_{2} $ \\
				 $ T $   &  Time block length \\
				 $ \omega $ & System bandwidth \\      
				 $ f_{k, \max } $  &  \tabincell{l}{Local computing capability of the  $ k $-th WD}  \\                           
			    $ C_{k} $ &  Number of CPU cycles required to process an input bit  \\
			    $ \kappa $ &  Energy efficiency coefficient \\
			    $ \eta  $ & Energy conversion efficiency \\
				 $ \mu $  & \tabincell{r}{Power consumption of each reflection element} \\
				\bottomrule
			\end{tabular}
		}
	\end{table}

	\emph{Notations:} In this paper, boldface lower- and upper-case letters represent vectors and matrixs, respectively. $[\cdot]^*,[\cdot]^T$, and  $[\cdot]^H$  denote the conjugate, transpose, and conjugate-transpose operations, respectively. $ \mathbb{E}\left[\cdot\right] $ is the expectation operator. $ {\rm diag}\left(\cdot \right) $ is the diagonal operation. $ \operatorname{tr}\left(\cdot \right) $, $  \operatorname{rank}(\cdot) $, and $ [\cdot]^{-1} $ denote the trace, rank, and inverse operations, respectively. $ \mathbb{C}^{M\times N} $ refers to a space of $ M\times N $ complex matrices; $ j $ is the imaginary unit. $ \mathbf{I}_{N} $ denotes an $ N\times N $ identity matrix. $ \mathbf{1}_{L} $ represents a $ L $-length vector with all elements are 1. $ \|\cdot\| $ represents the Euclidean norm of its argument. $ \otimes $ is the Kronecker product. $ \operatorname{Re}\left\lbrace \cdot \right\rbrace  $ denotes the real part of $ \cdot $.
	
	\section{System Model and Problem Formulation}
	In this section, we first introduce the system model, including phase shift model, uplink task offloading model, downlink energy transfer model, and computing model. Then, we formulate the computing rate maximization problem under several practical constraints. 
	\begin{figure}[htbp]
		\begin{minipage}[t]{0.42\textwidth}
			\centering
			\includegraphics[width=1.1\textwidth]{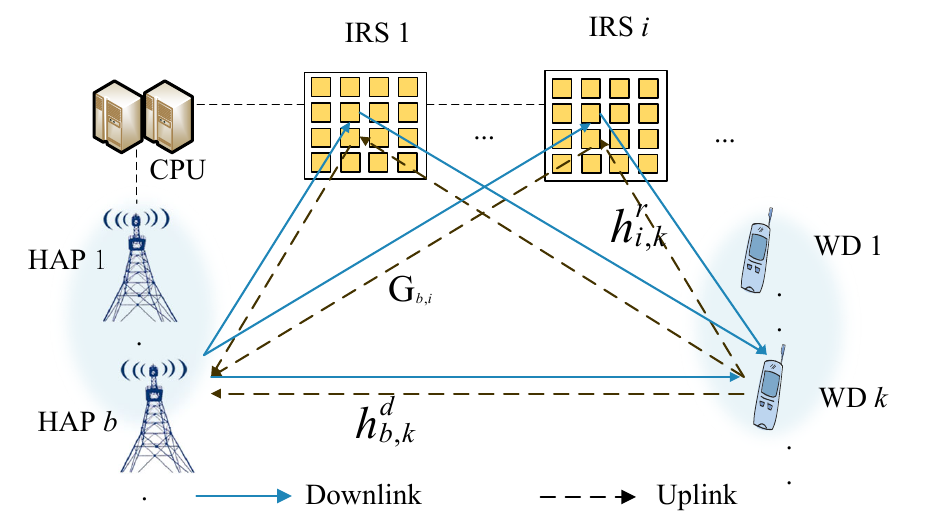}
			\caption{ IRS-aided WP-MEC networks.}
		\end{minipage}
	\end{figure}
	\begin{figure}[htbp]
		\begin{minipage}[t]{0.42\textwidth}
			\centering
			\includegraphics[width=1.1\textwidth]{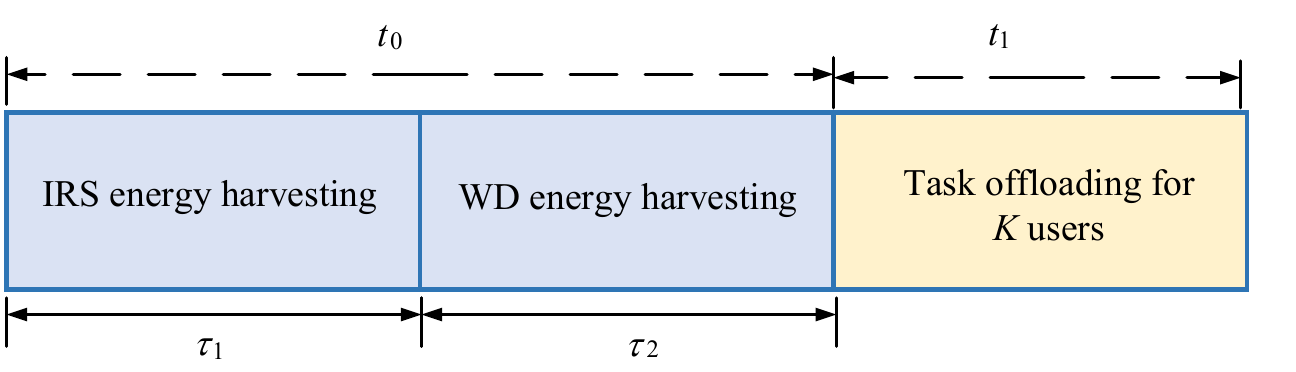}
			\caption{ An illustration of the harvest-then-computing protocol.}
		\end{minipage}
	\end{figure}
	\subsection{System Model}
	As illustrated in Fig. 1, we consider an IRS-assisted WP-MEC network, which comprises of $ B $ distributed HAPs indexed by $ \mathcal{B} \in\{1,\dots, B\} $, $ I $ IRSs indexed by $ \mathcal{I} \in\{1,\dots, I\} $, and $ K $ energy-constrained WDs indexed by $ \mathcal{K} \in\{1,\dots, K\} $. Each HAP is equipped with an edge server and all HAPs are interconnected to a central processing unit (CPU) through reliable backhaul links. Similar to \cite{ref25,ref26,ref27}, we assume that IRSs and WDs do not have any embedded energy supply available, and they need to collect the energy from the HAPs. After harvesting energy, IRSs can assist the downlink energy transfer and uplink task offloading, and WDs can process its computing task by local computing and task offloading. As shown in Fig. 2, we adopt the “harvest-then-computing” mechanism \cite{ref28, ref29}, and the system operates in a two-phase manner in each time block $ T $, which is assumed to be no larger than the tolerant latency of computing task. The number of antennas at the $ b $-th HAP, and the number of reflecting elements at the $ i $-th IRS are $ M_{b} $ and $ N_{i} $, respectively. For convenience, we set $ M_{b} $ and $ N_{i} $ equal  to $ M $ and $ N $ for all $ b \in \mathcal{B} $ and $ i \in \mathcal{I} $, respectively.

	The channel coefficients from the $ b $-th HAP to the $ k $-th WD, from the $ b $-th HAP to the $ i $-th IRS, and from the $ i $-th IRS to the $ k $-th WD are denoted by $ \mathbf{h}^{d}_{b,k} \in \mathbb{C}^{M\times 1} $ , $ \mathbf{G}_{b,i} \in \mathbb{C}^{M\times N} $ and $ \mathbf{h}^{r}_{i,k} \in \mathbb{C}^{N\times 1} $, respectively. Without loss of generality, we assume that all channels remain constant over each time block, and the channel state information (CSI) can be perfectly obtained by HAPs via the advanced channel estimation technique proposed in \cite{ref30}. The results with perfect CSI in this work serve as a theoretical performance upper bound for the practical system. The equivalent channel $ \mathbf{h}_{b,k} \in \mathbb{C}^{M\times 1} $ from the $ b $-th HAP to the $ k $-th WD can be expressed as
	\begin{equation}\label{Eq1}
		\mathbf{h}_{b,k}=\mathbf{h}^{d}_{b,k}+\sum_{i=1}^{I} \mathbf{G}_{b,i}\mathbf{\Theta}_{i} \mathbf{h}^{r}_{i,k},
	\end{equation}
	where $ \mathbf{\Theta}_{i} \in \mathbb{C}^{N\times N} $ denotes the phase shift matrix at the $ i $-th IRS, which is defined as
	\begin{equation} 
		\mathbf{\Theta}_{i} \triangleq {\rm diag}\left( \beta_{i,1} e^{j \theta_{i,1}},..., \beta_{i,N}e^{j \theta_{i,N}} \right), \forall i\in \mathcal{I}, 
	\end{equation}
	where $ \beta_{i,n} \in \left[ 0,1\right]  $ and $ \theta_{i,n} \in \left[0,2\pi  \right) $ denote the reflecting amplitude and phase shift at the $ n $-th reflecting element of the $ i $-th IRS. 

	\subsubsection{Phase Shift Model}
		\begin{figure}[thbp]
		\begin{minipage}[t]{0.42\textwidth}
			\centering
			\includegraphics[width=0.75\textwidth]{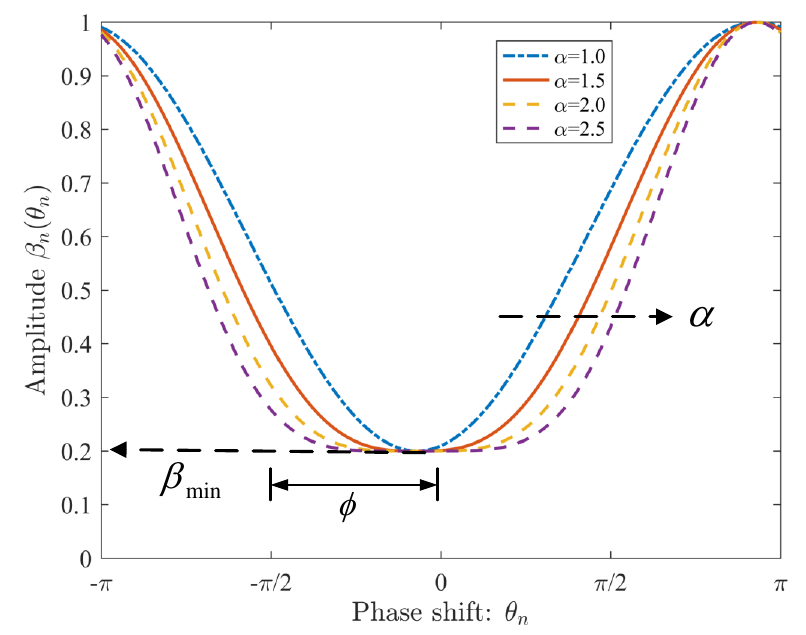}
			\caption{ The practical phase shift model with different parameters.}
		\end{minipage}
	\end{figure}
	We apply the practical phase shift model here, and it characterizes the fundamental relationship between the reflecting amplitude and the phase shift. Let $ v_{i,n}=\beta_{i,n}\left( \theta_{i,n}\right)e^{j\theta_{i,n}}  $,  $ \forall i \in \mathcal{I} $, $ \forall n \in \mathcal{N}  $, $ \beta_{i,n}\left( \theta_{i,n}\right) $ is given by \cite{ref10}
	\begin{equation}
		\beta_{i,n}\left(\theta_{i,n}\right)=\left(1-\beta_{\min }\right)\left(\frac{\sin \left(\theta_{i,n}-\phi\right)+1}{2}\right)^\alpha+\beta_{\min },
	\end{equation}
	where $ \beta_{\min } $, $ \phi $, and $ \alpha $ are the positive constants related to the circuit implementation. As depicted in Fig. 3, $ \beta_{\min } $ is the minimum amplitude, $ \phi $ is the horizontal distance between $ -\pi / 2 $ and the minimum amplitude $ \beta_{\min } $, and $ \alpha $ controls the steepness of the function curve. Note that  $ \beta_{\min }=1 $ corresponds to the ideal phase shift model with unit amplitude.
	
	\subsubsection{Uplink Task Offloading Model}
	Let $ \mathbf{s}=\left[ s_{1}, s_{2},...,s_{K} \right]^{T}  $, where $ s_{k} $ denotes the transmit signal of the $ k $-th WD and $\mathbb{E}\left\{\mathbf{s} \mathbf{s}^H\right\}=\mathbf{I}_K$. Thus, the received signal at the $ b $-th HAP for the $ k $-th WD can be expressed as
	\begin{equation}
		\hat{s}_{b,k}=\mathbf{u}_{b,k}^{H}\left[\sqrt{P_{k}} \sum_{j=1}^{K} \mathbf{h}_{b,j}s_{j}\! +\!\mathbf{n}_{b} \right],
	\end{equation}
	where $ \mathbf{u}_{b,k} $ denotes the MUD vector for the $ k $-th WD, $ P_{k} $ is the transmit power of the $ k $-th WD, and $ \mathbf{n}_{b}=\left[n_{b,1},n_{b,2},...,n_{b,M} \right] ^{T} $ is the received noise vector of the $ b $-th HAP where we assume $n_{b,m} \sim \mathcal{C} \mathcal{N}\left(0, \sigma^2\right)$ for all $ b \in \mathcal{B} $, and $ m \in \mathcal{M} $. Since there are $ B $ HAPs serving $ K $ WDs simultaneously, the overall recovered signal $ \hat{s}_{k} $ for the $ k $-th WD can be formulated as
	\begin{small}
		\begin{align}
			&\hat{s}_{k}=\sum_{b=1}^{B}\mathbf{u}_{b,k}^{H}\left[ \sum_{j=1}^{K} \sqrt{P_{k}} \left(\mathbf{h}^{d}_{b,j}\!+\!\sum_{i=1}^{I} \mathbf{G}_{b,i} \mathbf{\Theta}_{i} \mathbf{h}^{r}_{i,j}\right)s_{j}\! +\!\mathbf{n}_{b} \right]  \nonumber \\
			&\;\;\;\overset{\left( a \right) }{=}\mathbf{u}_{k}^{H}\left[ \sum_{j=1}^{K} \sqrt{P_{k}} \left(\mathbf{h}^{d}_{j}\!+\!\sum_{i=1}^{I} \mathbf{G}_{i} \mathbf{\Theta}_{i} \mathbf{h}^{r}_{i,j}\right)s_{j} +\mathbf{n} \right] \nonumber \\
			&\;\;\;\overset{\left( b \right) }{=}\mathbf{u}_{k}^{H}\left[ \sum_{j=1}^{K} \sqrt{P_{k}}\left(\mathbf{h}^{d}_{j}\!+\! \mathbf{G} \mathbf{\Theta} \mathbf{h}^{r}_{j}\right)s_{j} +\mathbf{n} \right] \\
			&\;\;\;\overset{\left( c \right) }{=}\mathbf{u}_{k}^{H}\left[ \sum_{j=1}^{K} \sqrt{P_{k}}\mathbf{h}_{j}s_{j} +\mathbf{n} \right], \nonumber
		\end{align}	
	\end{small} 
 $ \left( a \right) $ holds by defining $ \mathbf{u}_{k}=\left[\mathbf{u}_{1,k}^{T},..., \mathbf{u}_{B,k}^{T} \right]^{T}  $, $ \mathbf{h}^{d}_{k}=\left[ {\mathbf{h}^{d}_{1,k}}^{T},...,{\mathbf{h}^{d}_{B,k}}^{T} \right]^{T} $, $ \mathbf{G}_{i}=\left[\mathbf{G}_{1,i}^{T},...,\mathbf{G}_{B,i}^{T} \right]^{T}  $, and $ \mathbf{n}=\left[\mathbf{n}_{1}^{T},...,\mathbf{n}_{B}^{T} \right]^{T}  $, $ \left( b \right) $ holds by defining $ \mathbf{G}=\left[\mathbf{G}_{1}^{T},...,\mathbf{G}_{I}^{T} \right]^{T}  $,  $ \mathbf{\Theta}={\rm diag} \left(  \mathbf{\Theta}_{1},...,\mathbf{\Theta}_{I}\right)   $, and $ \mathbf{h}^{r}_{k}=\left[{\mathbf{h}^{r}_{1,k}}^{T},...,{\mathbf{h}^{r}_{I,k}}^{T} \right]^{T}$, and $ \left( c \right) $ holds according to $ \mathbf{h}_{k}=\mathbf{h}^{d}_{k}+ \mathbf{G} \mathbf{\Theta} \mathbf{h}^{r}_{k}$. Thus, the SINR of the $ k $-th WD can be expressed as
	\begin{equation}\label{Eq05}
		\gamma_{k}\!\left(\mathbf{P},\mathbf{U}, \mathbf{\Theta} \right)\!=\!\dfrac{P_{k}\left| \mathbf{u}_{k}^{H}\left(\mathbf{h}^{d}_{k}\!+\! \mathbf{G} \mathbf{\Theta} \mathbf{h}^{r}_{k}\right) \right|^{2} }{\sum_{j=1,j\ne k}^{K} P_{j} \left| \mathbf{u}_{k}^{H}\!\left(\mathbf{h}^{d}_{j}\!+\! \mathbf{G} \mathbf{\Theta} \mathbf{h}^{r}_{j}\right) \right|^{2}\!\!+\!\sigma^{2} },
	\end{equation}
  where $ \mathbf{P}=\left[ P_{1},..., P_{K} \right] $, $ \mathbf{U}=\left[ \mathbf{u}_{1}^{T}, ..., \mathbf{u}_{K}^{T}\right]^{T}  $. Thereby, the offloading rate of the $ k $-th WD is given by
	\begin{eqnarray}\label{Eq06}
		R_{k}\left(\mathbf{P}, \mathbf{U}, \mathbf{\Theta} \right)=\omega \log_2\left(1+\gamma_{k}\left(\mathbf{P}, \mathbf{U}, \mathbf{\Theta} \right)\right), 
	\end{eqnarray}
	where $ \omega $ is the bandwidth.

	\subsubsection{Downlink Energy Transfer Model }
	At the beginning of each time block as shown in Fig. 2, HAPs first generate energy beamforming to charge IRSs or WDs. We further divide the energy transfer duration $ t_{0} $ into two sub-slots $ \tau_{1} $ and $ \tau_{2} $, which satisfies $ t_{0}=\tau_{1}+\tau_{2} $.
	
	In the first sub-slot $ \tau_{1} $, IRSs operate in the energy harvesting mode for collecting the radio-frequency energy from HAPs. Note that the harvested direct current power is generally nonlinear with respect to the received radio frequency power. The nonlinear conversion largely depends on the input power level and the transmit waveform and there have been a plethora of recent works on analytic nonlinear energy harvesting (EH) model. However, there is still a lack of a generic EH model that captures all practical issues \cite{ref31}. Therefore, we consider a linear EH model here, which has been commonly adopted in the WPT literature \cite{ref32, ref33, ref34, ref35}. Let $ \mathbf{e} \in \mathbb{C}^{BM\times 1} $ denote the downlink energy-bearing transmit signal, which is assumed as a random signal with its power spectral density satisfying certain regulations on radio-frequency radiation \cite{ref36}. Let $ \mathbf{W} \triangleq \mathbb{E}\left[\boldsymbol{\mathbf{e} \mathbf{e}^ { H }}\right] \succeq \mathbf{0}$ denote the energy transmit covariance matrix and the transmit power is given by $\mathbb{E}\left[\|\mathbf{e}\|^2\right]=\operatorname{tr}(\mathbf{W})$. In general, HAPs can apply multiple beams to deliver wireless energy, i.e., $ \mathbf{W} $ can be of any rank. Assuming $r=\operatorname{rank}(\mathbf{W}) \leq BM$, then a total of $ r $ energy beams can be obtained via the eigenvalue decomposition (EVD) of $ \mathbf{W} $ \cite{ref36}. Since the noise power is much lower than the received power transmitted from HAPs, the energy harvested from the noise can be ignored. Hence, the harvested energy of the $ i $-th IRS can be expressed as 
	\begin{equation}
		E_{IRS}\left(i, \tau_{1}, \mathbf{W}\right)= \tau_{1} \eta \operatorname{tr} \left( \mathbf{G}^{'}_{i}  \mathbf{W} \right),
	\end{equation}
	where  $ \eta $ denotes the linear energy conversion efficiency, and $ \mathbf{G}^{'}_{i}  \triangleq \mathbf{G}_{i}  \mathbf{G}_{i}^{H}$. 
	
	In the second sub-slot $ \tau_{2} $, HAPs charge WDs with  the help of IRSs. Let $ \mathbf{x} \in \mathbb{C}^{BM\times 1} $ denote the energy-bearing transmit signal. $ \mathbf{Q} \triangleq \mathbb{E}\left[\boldsymbol{\mathbf{x} \mathbf{x}^ { H }}\right] \succeq \mathbf{0}$ denotes the energy transmit covariance matrix and the transmit power is given by $\mathbb{E}\left[\|\mathbf{x}\|^2\right]=\operatorname{tr}(\mathbf{Q})$. 
	The energy harvested by the $ k $-th device is given by
	\begin{equation}
	E_{k}\left(\tau_{2},\mathbf{Q}, \mathbf{v}^{E} \right)=\tau_{2}\eta\operatorname{tr}\left(\mathbf{Q}\mathbf{H}_{k} \right),
\end{equation}
	where $\mathbf{H}_k \triangleq \mathbf{h}_k \mathbf{h}_k^H$, and $\mathbf{v}^{E}=\mathbf{\Theta}\mathbf{1}_{IN}$ denotes the IRS passive beamforming at the second energy harvesting sub-slot. Additionally, WDs can also collect the wireless energy at the first sub-slot, but the energy is very weak since HAPs mainly formulate the high gain energy beamforming towards to RISs, thus can be omitted. 
	
	\subsubsection{Computing Model}
	In the second phase, WDs utilize the harvested energy to perform local computing and task offloading with the duration of $ t_{1} $, which satisfies $ t_{1}=T-t_{0} $.
    \paragraph{Local Computing}
	Upon denoting the CPU cycles required to process a single bit input data and the computing speed of the $ k $-th WD by $ c_{k} $ and $ f_{k} $, respectively, we formulate the local computing data size and energy consumption of the $ k $-th WD as \cite{ref37} 
	\begin{subequations}
		\begin{align}
			& D_{k, loc}=\frac{f_k t_{1}}{c_k}, \\
			& E_{k, loc}=\kappa f_k^2 t_{1},
		\end{align}
	\end{subequations}
	where $ \kappa $ is computation energy efficiency coefficient, which depends on the hardware architecture \cite{ref38, ref39}.
	\paragraph{Task Offloading}
	Apart from local computing, WDs can also choose to offload a part of its computing task to the edge server. Combining formula $ \left( \ref{Eq06} \right) $, the offloading data size can be expressed as
	\begin{equation}
		D_{k, off}= t_1 R_{k}\left( \mathbf{P}, \mathbf{U}, \mathbf{v}^{I} \right),
	\end{equation}  
	where $\mathbf{v}^{I}=\mathbf{\Theta}\mathbf{1}_{IN}$ is the IRS passive beamforming at the data computing phase. Since the edge server has powerful computing capability compared with WDs and the size of computing result is usually small, thus we ignore the delay of edge computing and result reture as in \cite{ref40, ref41, ref42}.
	
	\subsection{Problem Formulation}
	In this paper, we aim to maximize the total computation bits of all WDs by jointly optimizing the downlink/uplink passive beamforming $ \left\lbrace \mathbf{v}^{E}, \mathbf{v}^{I} \right\rbrace  $, downlink energy beamforming $ \left\lbrace \mathbf{W}, \mathbf{Q} \right\rbrace $, uplink MUD vector $ \mathbf{U} $, offloading power $ \mathbf{P}$ and local CPU-cycle frequency $ \mathbf{f}=\left[f_{1},...,f_{K} \right]  $ at WDs, and the time slot allocation $ \mathbf{t}=[\tau_{1},\tau_{2},t_1] $, and the joint optimization problem is formulated as 
	\begin{subequations}\label{OptA}
		\begin{align}
			&\mathcal{P} 0: \underset{\left\lbrace \mathbf{x}_{0}\right\rbrace }{\operatorname{maximize}} \quad t_{1} \sum_{k=1}^{K} \omega \log_{2} \left(1+\gamma_{k} \right)+ \sum_{k=1}^{K} \frac{f_{k}t_{1}}{C_{k}}   \\
			&\text { s.t. }  \tau_{1}+\tau_{2}+t_{1}\leq T, \label{OptA1} \\
			&  0 \leq f_{k} \leq f_{k, \max }, \forall k \in \mathcal{K}, \label{OptA2} \\
			&   v_{i,n}=\beta_{i,n}\left(\theta_{i,n}\right) e^{\jmath \theta_{i,n}}, \forall i \in \mathcal{I}, \forall n \in \mathcal{N}, \label{OptA3}  \\
			& -\pi \leq \theta_{i,n} \leq \pi, \forall i \in \mathcal{I}, \forall n \in \mathcal{N}, \label{OptA4}  \\ 
			& \operatorname{tr}\left( \mathbf{W} \mathbf{D}_{b}\right) \leq P_{\max}, \forall b \in \mathcal{B}, \label{OptA5} \\
			& \operatorname{tr} \left( \mathbf{Q} \mathbf{D}_{b}\right) \leq P_{\max}, \forall b \in \mathcal{B},\label{OptA6} \\ 
			& \left\| \mathbf{u}_{b,k} \right\|^{2} \leq 1, \forall b \in \mathcal{B}, \forall k \in \mathcal{K}, \label{OptA7}  \\
			& \left( T-\tau_{1} \right) N\mu \leq E_{IRS}\left(i, \tau_{1}, \mathbf{W}\right), \forall i \in \mathcal{I}, \label{OptA8} \\
			&	t_{1}\left[\kappa f_{k}^{2}\!+\! \left(P_{k}\!+\!P_{c}\right)\right] \! \leq \! E_{k}\left(\tau_{2},\mathbf{Q}, \mathbf{v}^{E} \right)\!, \forall k \! \in \mathcal{K}, \label{OptA9} 
		\end{align}
	\end{subequations}
	where $ \mathbf{D}_{b}=\mathbf{C}_{b} \otimes \mathbf{I}_{M}$, $ \mathbf{C}_{b} \in \mathbb{C}^{B\times B} $ is a matrix with the element at the $ b $-th row and the $ b $-th column being 1, and the other elements being 0s. $ \mu $ is the power consumption of each IRS reflecting element \cite{ref43, ref44}. $ P_{\max} $ and $ P_{c} $ denote the maximum transmit power of each HAP and the constant circuit power. The optimization variables $\left\lbrace \mathbf{x}_{0} \right\rbrace =\left\lbrace \mathbf{W},  \mathbf{Q}, \mathbf{v}^{E}, \left\lbrace \theta_{i,n}^{E} \right\rbrace,   \mathbf{U},  \mathbf{v}^{I}, \left\lbrace \theta_{i,n}^{I} \right\rbrace, \mathbf{P}, \mathbf{f}, \mathbf{t} \right\rbrace$.
   $ \left( \ref{OptA1} \right)  $ indicates the time constraint, $ \left( \ref{OptA2} \right)  $ is the local computing capability constraint, $ \left(\ref{OptA3} \right)  $ gives the reflecting coefficient constraint of IRSs, $ \left(\ref{OptA4} \right)  $ restricts the range of the phase shift, $ \left(\ref{OptA5} \right)  $ and $ \left(\ref{OptA6} \right)  $ represent the downlink transmit power constraints, $ \left(\ref{OptA7} \right) $ is the unit-norm detection vector constraint, and $ \left(\ref{OptA8} \right)  $ and $ \left(\ref{OptA9} \right)  $ imply that the energy consumption of IRSs and WDs should not exceed their harvested energy. It is obvious that $ \mathcal{P} 0 $ is a nonconvex optimization problem and thus intractable. Next, we propose an effective scheme to deal with it. 
	
	\section{Solution of the Optimization Problem}
	In the first sub-slot $ \tau_{1} $, HAPs transmit energy beamforming to charge IRSs, and then IRSs assist the downlink energy transfer in the second sub-slot $ \tau_2 $ and uplink task offloading in the second phase $ t_1 $. From $ \mathcal{P} 0 $, one can observe that the computing data size increases with $ t_1 $, and thus we can decrease $ \tau_1 $ and increase $ \tau_2 $/$ t_1 $ to enhance the harvested energy by WDs or offloading time so as to increase computation bits. By combining with $ \left( \ref{OptA8} \right)  $, we can directly obtain the optimal $ \tau_1 $ for the $ i $-th IRS as 
	 \begin{equation}\label{Eq16}
		\tau_{1,i}^\star =\frac{N \mu T}{N \mu+\eta \operatorname{tr} \left( \mathbf{G}^{'}_{i} \mathbf{W} \right) }, \forall i \in \mathcal{I}.
	\end{equation}
	To satisfy the energy requirement for each IRS and simultaneously minimize the sub-slot $ \tau_1 $, we formulate the following optimization problem. 
		\begin{subequations} \label{OptEE}
		\begin{align}
			& \mathcal{P} 1: \underset{\mathbf{W}, \tau_{1}}{\operatorname{min}} \quad \tau_{1}\\
			\text { s.t. }	& \frac{N \mu T}{N \mu+\eta \operatorname{tr} \left( \mathbf{G}^{'}_{i} \mathbf{W} \right) } \leq \tau_{1}, \forall i \in \mathcal{I}, \label{OptEE1} \\
			& \operatorname{tr}\left(\mathbf{W}\mathbf{D}_{b} \right)  \leq P_{\max}, \forall b \in \mathcal{B}, \label{OptEE2}\\
			& \mathbf{W} \succeq \mathbf{0} \label{OptEE3}.
		\end{align}
	\end{subequations}
	It is obvious that $ \mathcal{P} 1 $ is a non-convex optimization problem, and we reformulate $ \left( \ref{OptEE1} \right)  $ as $ N \mu T \leq \tau_{1} \left( {N \mu+\eta \operatorname{tr} \left( \mathbf{G}^{'}_{i} \mathbf{W} \right) } \right) $. Then, with the help of a bisection search over $ \tau_{1} $, $ \mathcal{P} 1 $ can be equivalently transformed into the following convex optimization problem, given as
	\begin{subequations} \label{OptEEE}
	\begin{align}
		& \mathcal{P} 1^{'}: {\operatorname{Find}} \quad \mathbf{W}\\
		\text { s.t. }	& N \mu T \leq \tau_{1}^{l} \left( {N \mu+\eta \operatorname{tr} \left( \mathbf{G}^{'}_{i} \mathbf{W} \right) } \right) , \forall i \in \mathcal{I} \\
		& \operatorname{tr}\left(\mathbf{W}\mathbf{D}_{b} \right)  \leq P_{\max}, \forall b \in \mathcal{B},\\
		& \mathbf{W} \succeq \mathbf{0},
	\end{align}
    \end{subequations}
   where $ \tau_{1}^{l} $ is the value of $ \tau_{1} $ at the $ l $-th iteration.

    After obtaining the optimal $ \tau_{1} $ and $ \mathbf{W} $, the original $ \mathcal{P} 0 $ can be equivalently transformed into the following problem
     \begin{subequations} \label{OptB}
    	\begin{align}
    		\mathcal{P} 2: & \underset{\left\lbrace \mathbf{x}_{1}\right\rbrace}{\operatorname{maximize}} \quad {t}_{1} \sum_{k=1}^{K} \omega \log_{2} \left(1+\gamma_{k} \right)+ \sum_{k=1}^{K} \frac{f_{k}{t}_{1}}{C_{k}}  \\
    		\text { s.t. } & \tau_{2}+t_{1}\leq T-\tau_{1} \\
    		&\left(\ref{OptA2} \right) - \left(\ref{OptA4} \right), \left(\ref{OptA6} \right), \left(\ref{OptA7} \right), \left(\ref{OptA9} \right), \label{OptB1} 
    	\end{align}
    \end{subequations}
    where $\left\lbrace \mathbf{x}_{1} \right\rbrace =\left\lbrace  \mathbf{Q}, \mathbf{v}^{E}, \left\lbrace \theta_{i,n}^{E} \right\rbrace, \mathbf{U},  \mathbf{v}^{I}, \left\lbrace \theta_{i,n}^{I} \right\rbrace, \mathbf{P}, \mathbf{f}, \tau_{2} \right\rbrace$. One can observe that it is still difficult to directly solve $\mathcal{P} 2$. In the rest of this section, we will propose an effective algorithm to solve it.
    
   Note that $ \left\lbrace \mathbf{Q}, \mathbf{v}^{E}, \left\lbrace \theta_{i,n}^{E} \right\rbrace \right\rbrace  $ are related to the WPT setting of the second sub-slot, and $ \left\lbrace \mathbf{U},  \mathbf{v}^{I}, \left\lbrace \theta_{i,n}^{I} \right\rbrace,   \mathbf{P}, \mathbf{f} \right\rbrace $ are related to the computing setting of the second phase. Therefore, if we fix $ \tau_{2} $, $ \mathcal{P} 2 $ can be divided into the following two independent sub-problems, i.e., $ \mathcal{P} 3 $ and $ \mathcal{P} 4 $.
    \begin{subequations} \label{OptC}
    	\begin{align}
    		&\mathcal{P} 3: Find \left\lbrace \mathbf{Q}, \mathbf{v}^{E}, \left\lbrace \theta_{i,n}^{E} \right\rbrace \right\rbrace\\
    		&\quad \text { s.t. }	\left(\ref{OptA3} \right), \left(\ref{OptA4} \right),\left(\ref{OptA6} \right), \label{OptC1} \\
    		&\quad t_{1}\left[\kappa f_{k}^{2}\!+\! \left(P_{k}\!+\!P_{c}\right)\right] \! \leq \! \tau_{2}\eta\operatorname{tr}\left(\mathbf{Q}\mathbf{H}_{k} \right), \forall k \in \mathcal{K}, \label{OptC3}  
    	\end{align}
    \end{subequations}
   where $ t_{1}=T-\tau_{1}-\tau_{2} $.
   \begin{subequations}\label{OptL} 
    	\begin{align}
    		\mathcal{P} 4: & \underset{ \left\lbrace \mathbf{x}_{2} \right\rbrace }{\operatorname{max}} \quad  \sum_{k=1}^{K} \omega \log_{2} \left(1+\gamma_{k} \right)+ \sum_{k=1}^{K}\frac{f_{k}}{C_{k}}  \label{OptL1} \\
    		\text { s.t. } & \left( \ref{OptA2}\right)-\left( \ref{OptA4}\right), \left( \ref{OptA7}\right),  \\
    		& t_{1}\left[\kappa f_{k}^{2}\!+\!\left(P_{k}\!+\!P_{c}\right)\right] \! \leq \! \tau_{2}\eta\operatorname{tr}\left(\mathbf{Q}\mathbf{H}_{k} \right), \forall k \in \mathcal{K}, \label{OptL6}
    	\end{align}
    \end{subequations}
	where $ \left\lbrace \mathbf{x}_{2} \right\rbrace = \left\lbrace \mathbf{U}, \mathbf{v}^{I}\!, \left\lbrace \theta_{i,n}^{I} \right\rbrace,  \mathbf{P}, \mathbf{f} \right\rbrace  $. 
	
	Here, we need to solve $ \mathcal{P} 3 $ first and then solve $ \mathcal{P} 4 $ according to the obtained $ \left\lbrace \mathbf{Q}, \mathbf{v}^{E}, \left\lbrace \theta_{i,n}^{E} \right\rbrace \right\rbrace  $ based on $ \mathcal{P} 3 $. Meanwhile, we can apply one dimensional search method to obtain the optimal $ \tau_2 $. Next, we only need to solve $ \mathcal{P} 3 $ and $ \mathcal{P} 4 $ in sequence.  
    
	\subsection{Solution of $ \mathcal{P}_{3} $}
  In fact, there may be many feasible solutions for $\mathcal{P} 3$, and it seems difficult to decide which one is the optimal. However, from $ \left( \ref{OptL6} \right)  $ of $ \mathcal{P} 4$, one can observe that a larger $ \operatorname{tr}\left(\mathbf{Q}\mathbf{H}_{k} \right) $ can obtain a larger objective function for $ \mathcal{P} 4$. Since $ \operatorname{tr}\left(\mathbf{Q}\mathbf{H}_{k} \right) $ is related to $ \mathcal{P} 3$, we can select  one of feasible solutions of $ \mathcal{P} 4$  that owns the maximum $ \operatorname{tr}\left(\mathbf{Q}\mathbf{H}_{k} \right) $. Based on the above analysis, we can transform $ \mathcal{P} 3$ into the following optimization problem as 
    \begin{subequations} \label{OptDD}
    	\begin{align}
    		\mathcal{P} 3a:  &\underset{ \mathbf{Q}, \mathbf{v}^{E}, \left\lbrace \theta_{i,n}^{E} \right\rbrace}{\operatorname{max}}  \sum_{k=1}^{K} \operatorname{tr}\left(\mathbf{Q}\mathbf{H_{k}} \right) \\
    	    \text { s.t. } & \left(\ref{OptA3} \right), \left(\ref{OptA4} \right),\left(\ref{OptA6} \right), \label{OptDD1} \\
    		& \mathbf{Q} \succeq \mathbf{0}, \label{OptDD2}\\
    		& \operatorname{rank}(\mathbf{Q})=1. \label{OptDD3} 
    	\end{align}
    \end{subequations}
   To facilitate the solution, we transform  $\mathcal{P} 3a$ to its equivalent problem, by introducing  auxiliary variable $ \left\lbrace \zeta_{k} \right\rbrace  $ as 
	\begin{subequations} \label{OptD}
		\begin{align}
			\mathcal{P} 3a^{\prime}:&  \underset{ \mathbf{Q}, \mathbf{v}^{E}, \left\lbrace \theta_{i,n}^{E} \right\rbrace, \left\lbrace \zeta_{k} \right\rbrace }{\operatorname{max}}  \sum_{k=1}^{K} \zeta_{k} \\
			\text { s.t. }&	\left(\ref{OptA3} \right), \left(\ref{OptA4} \right),\left(\ref{OptA6} \right), \left( \ref{OptDD2} \right), \left( \ref{OptDD3} \right), \label{OptD1} \\
			&\operatorname{tr}\left(\mathbf{Q}\mathbf{H_{k}} \right) \geq \zeta_{k},  \forall k \in \mathcal{K}. \label{OptD3}  
		\end{align}
	\end{subequations}
 Due to the non-convex constraint $ \left( \ref{OptA3} \right)  $, it is difficult to solve $ \mathcal{P} 3a^{\prime} $. To proceed it, we propose a penalty-based method by adding a penalty term to the OF, which can be reformulated as
\begin{subequations}\label{OptH} 
	\begin{align}
		&\mathcal{P} 3a^{\prime}-1: \notag \\
		& \underset{\mathbf{Q}, \mathbf{v}^{E}, \left\lbrace \theta_{i,n}^{E} \right\rbrace, \left\lbrace \zeta_{k} \right\rbrace }{\operatorname{min}} \!-\!\sum_{k=1}^{K} \zeta_{k}\!+\!\iota_{1} \sum_{n=1}^{N} \left| v_{i,n} \!-\! \beta_{i,n}\left(\theta_{i,n}\right) e^{\jmath \theta_{i,n}} \right|^{2} \label{OptH1}\\
		&\text { s.t. }  \left(\ref{OptA4} \right),\left(\ref{OptA6} \right), \left( \ref{OptDD2} \right), \left( \ref{OptDD3} \right), \left( \ref{OptD3} \right),  
	\end{align}
\end{subequations}
where $ \iota_{1}>0 $ is the penalty parameter that imposes a cost for the constraint violation of $ \left( \ref{OptA3} \right)  $. We propose a two-layer iterative algorithm, where the inner layer solves the penalized optimization problem $ \mathcal{P} 3a^{\prime}-1 $ by adopting the  block coordinate descent (BCD) method while the outer layer updates $ \iota_{1} $, until the convergence is achieved. Specifically, the optimization variables in $ \mathcal{P} 3a^{\prime}-1 $ can be partitioned into three blocks, i.e., $ \mathbf{Q} $, $\mathbf{v}^{E}$ and $ \left\lbrace \theta_{i,n}^{E} \right\rbrace  $, and each of them is alternately optimized in one iteration with the other two blocks fixed, until the convergence is reached.

Note that the initial value of $ \iota_{1} $ should be set to a sufficiently small number, even though this point may be infeasible for $ \left(\ref{OptD} \right)  $. We can maximize the OF, i.e., $ \sum_{k=1}^{K} \zeta_{k} $, by gradually increasing $ \iota_{1} $ by a factor of $\varrho>1$. When $ \iota_{1} $ is sufficiently large, we can obtain a solution that satisfies all the constraints in $ \left( \ref{OptD} \right)  $ within a predefined accuracy.  

Next, we provide the details for solving $ \mathcal{P} 3a^{\prime}-1 $.

\subsubsection{Optimizing energy transmit covariance matrix $ \mathbf{Q} $ while fixing the passive beamforming $ \mathbf{v}^{E} $ and IRS phase shift $ \left\lbrace \theta_{i,n}^{E} \right\rbrace $}    
   For given $ \left\lbrace \mathbf{v}^{E}, \left\lbrace \theta_{i,n}^{E} \right\rbrace \right\rbrace $, $ \mathcal{P} 3a^{\prime} $ can be simplified as
  \begin{subequations} \label{OptF}
		\begin{align}
			&\mathcal{P} 3a^{\prime}-1E1:  \underset{\mathbf{Q}, \left\lbrace \zeta_{k} \right\rbrace}{\operatorname{max}}  \sum_{k=1}^{K} \zeta_{k} \\
			&\text { s.t. }  \left(\ref{OptA6} \right), \left( \ref{OptDD2}\right), \left( \ref{OptDD3} \right), \left( \ref{OptD3} \right).   	 
		\end{align}
  \end{subequations}   
For the rank-one constraint $ \left( \ref{OptDD3} \right)  $, it is equivalent to $ \operatorname{tr}\left( \mathbf{Q} \right)-\left\| \mathbf{Q} \right\|_{2}=0 $ \cite{ref70}, where $ \left\| \mathbf{Q} \right\|_{2} $ denotes the spectral norm. For a positive semidefinite matrix $ \mathbf{Q} $, there always exists $ \operatorname{tr}\left( \mathbf{Q} \right)-\left\| \mathbf{Q} \right\|_{2} \geq 0 $, since $ \left\| \mathbf{Q} \right\|_{2} $ is the largest singular value while $ \operatorname{tr}\left( \mathbf{Q} \right) $ equals to the sum of all singular value. Thus, $ \operatorname{tr}\left( \mathbf{Q} \right)-\left\| \mathbf{Q} \right\|_{2}=0 $ can be transformed into the difference of convex functions as
\begin{equation}\label{EQ21}
	\operatorname{Tr}(\mathbf{Q})-\|\mathbf{Q}\|_2 \leq 0.
\end{equation} 
Then, $ \mathcal{P} 3a^{\prime}-1E1 $ can be transformed into
     \begin{subequations} \label{OptF}
   	\begin{align}
   		&\mathcal{P} 3a^{\prime}-1E1^{\prime}:  \underset{\mathbf{Q}, \left\lbrace \zeta_{k} \right\rbrace}{\operatorname{max}}  \sum_{k=1}^{K} \zeta_{k} \\
   		&\text { s.t. }  \left(\ref{OptA6} \right), \left( \ref{OptDD2}\right), \left( \ref{OptD3} \right), \left( \ref{EQ21} \right).   	 
   	\end{align}
   \end{subequations}   
   It is obvious that $ \mathcal{P} 3a^{\prime}-1E1^{\prime} $ is a convex optimization problem, and thus it can be solved by using the standard convex optimization tool.
	
	\subsubsection{Optimizing passive beamforming $ \mathbf{v}^{E} $ while fixing energy transmit covariance matrix $ \mathbf{Q} $ and IRS phase shift $ \left\lbrace \theta_{i,n}^{E} \right\rbrace $} Firstly, we obtain the energy beamforming vector $ \mathbf{x} $ by singular value decomposition (SVD). Then, for convenience, we define $\mathbf{G} \mathbf{\Theta} \mathbf{h}_{r,k}=\boldsymbol{\Phi}_{ k}^H \boldsymbol{v}$, $ \mathbf{c}_{k}=\boldsymbol{\Phi}_{ k}\mathbf{x} $, $ d_{k}=\mathbf{h}_{d,k}^{H}\mathbf{x} $, $ \forall k \in \mathcal{K} $, and the left side of constraint $ \left( \ref{OptD3} \right)  $ can be rewritten as
	\begin{subequations}
		\begin{align}
			&\operatorname{tr}\left(\mathbf{Q}\mathbf{H_{k}} \right)=
			\mathcal{F}_{k}\!\left(\mathbf{v}^{E} \right) \notag  \\
			&=\left( \mathbf{v}^{E}\right) ^{H} \boldsymbol{C}_{k} \mathbf{v}^{E}\!+\!2 \operatorname{Re}\left\{\left( \mathbf{v}^{E}\right) ^{H} \mathbf{u}_{k}\right\}\!+\left| d_{k} \right|^{2},
		\end{align}
	\end{subequations}
	where $ \mathbf{C}_{k}=\mathbf{c}_{k}\mathbf{c}_{k}^{H} $ and $ \mathbf{u}_{k}=\mathbf{c}_{k}d_{k}^{H} $. Thus, $ \mathcal{P} 3a^{\prime}-1 $ can be simplified as
\begin{subequations}\label{OptHH} 
	\begin{align}
		\mathcal{P} 3a^{\prime}-1 E 2: & \underset{\mathbf{v}^{E},  \left\lbrace \zeta_{k} \right\rbrace }{\operatorname{min}} \! -\!\sum_{k=1}^{K} \zeta_{k}\!+\!\iota_{1} \left\|  \mathbf{v}^{E} - \mathbf{a}^{E} \right\| ^{2} \label{OptHH1}\\
		&\text { s.t. } \mathcal{F}_{k}\left(\mathbf{v}^{E} \right) \geq \zeta_{k},  \forall k \in \mathcal{K}, \label{OptHH2}
	\end{align}
\end{subequations}
 where $ \mathbf{a}^{E}=\left[ \beta_{1,1}\left(\theta^{E}_{1,1}\right) e^{\jmath \theta^{E}_{1,1}},..., \beta_{IN}\left(\theta^{E}_{IN}\right) e^{\jmath \theta^{E}_{IN}}  \right]^{T}  $. For the non-convex constraint $ \left(\ref{OptHH2} \right)  $, we apply the SCA technique to deal with it. The left hand side of $ \left(\ref{OptHH2} \right)  $ is lower bounded by its first-order Taylor expansion at any given point. Thus, at the local point of $ \left( \mathbf{v}^{E}\right) ^{\left( l-1\right) } $, we have
	\begin{equation}
		\begin{aligned}
			&\mathcal{F}_{k} \left(\mathbf{v}^{E}\right) \geq\\
			&\left( \mathbf{v}^{E}\right) ^{(l-1)^{H}} \mathbf{C}_{k} \left( \mathbf{v}^{E}\right) ^{(l-1)}\!+\!2 \operatorname{Re}\left\{\left( \mathbf{v}^{E}\right) ^{(l-1)^{H}} \mathbf{u}_{k}\right\}\!+\!\left|d_{k}\right|^{2} \\
			&+2\left(\mathbf{C}_{k}^{H} \left( \mathbf{v}^{E}\right) ^{(l-1)}+\mathbf{u}_{k}\right)^{H}\left(\left( \mathbf{v}^{E}\right) -\left( \mathbf{v}^{E}\right) ^{(l-1)}\right)\\
			&\triangleq \mathcal{F}_{k}^{\mathrm{low}}\left(\mathbf{v}^{E},\left( \mathbf{v}^{E}\right) ^{\left(l-1 \right) } \right).
		\end{aligned}
	\end{equation} 
  $ \mathcal{P} 3a^{\prime}-1 E 2  $ can be transformed into
	\begin{subequations}\label{OptI} 
		\begin{align}
			&\mathcal{P} 3a^{\prime}-1 E 2^{\prime}: \underset{ \mathbf{v}^{E}, \left\lbrace \zeta_{k} \right\rbrace }{\operatorname{min}} -\sum_{k=1}^{K} \zeta_{k} + \iota_{1}  \left\|  \mathbf{v}^{E} - \mathbf{a}^{E} \right\| ^{2}  \label{OptI1} \\
			&\text { s.t. }  \left| v_{i, n}^{E} \right| \leq 1, \forall i \in \mathcal{I}, \forall n \in \mathcal{N}, \label{OptI2} \\
			& \quad \quad \mathcal{F}_{k}^{\mathrm{low}}\left(\mathbf{v}^{E},\left( \mathbf{v}^{E}\right) ^{\left(l-1 \right) } \right)  \geq \zeta_{k}, \label{OptI3} \forall k \in \mathcal{K}.
		\end{align}
	\end{subequations}
	 A locally optimal solution can be approached by successively updating $ {\left( \mathbf{v}^{E}\right) }^{ \left( l-1 \right)  } $ based on the solution obtained from the previous iteration, until the objective value converges.

\subsubsection{Optimizing IRS phase shift $ \left\lbrace \theta_{i,n}^{E} \right\rbrace $ while fixing energy transmit covariance matrix $ \mathbf{Q} $ and  passive beamforming $ \mathbf{v}^{E} $} For given $ \mathbf{v}^{E} $, $ \mathcal{P} 3a^{\prime}-1 $ can be simplified as
	\begin{subequations}\label{OptJ} 
		\begin{align}
			&\mathcal{P} 3a^{\prime}-1 E 3: \notag \\
			& \quad \quad  \underset{ \left\lbrace \theta_{i,n}^{E} \right\rbrace}{\operatorname{min}} \sum_{i=1}^{I} \sum_{n=1}^{N}  \left| v_{i,n}^{E} - \beta_{i,n}\left(\theta_{i,n}^{E}\right) e^{\jmath \theta_{i,n}^{E}} \right|^{2}  \label{OptJ1}  \\
			&\quad \quad \text { s.t. } -\pi \leq \theta_{i,n} \leq \pi, \forall i \in \mathcal{I}, \forall n \in \mathcal{N}. \label{OptJ2}
		\end{align}
	\end{subequations}
	Due to the fact that $ \left\lbrace \theta_{i,n} \right\rbrace $ are fully separable in the OF, the solutions of $ \left\lbrace \theta_{i,n} \right\rbrace $ can be obtained by independently solving $ IN $ subproblems. By expanding $ \left| v_{i,n} - \beta_{i,n}\left(\theta_{i,n}\right) e^{\jmath \theta_{i,n}} \right|^{2} $ and omitting the constant terms, each subproblem can be formulated as
	\begin{subequations}\label{OptK}
		\begin{align}
			& \mathcal{P} 3a^{\prime}-1 E 3^{\prime}: \notag \\ 
			& \underset{\theta_{i,n}}{\operatorname{max}} 2 \beta_{i,n}\left(\theta_{i,n}\right)\left|v_{i,n}\right| \cos \left(\psi_{i,n}\!-\!\theta_{i,n}\right)\!-\!\beta_{i,n}^{2}\left(\theta_{i,n}\right) \label{OptK1} \\
			& \quad \text { s.t. } \left( \ref{OptJ2} \right),  \label{OptK2}
		\end{align}
	\end{subequations}
	where $\psi_{i,n}=\arg \left(v_{i,n}\right)$. As described in \cite{ref10}, the whole function is maximized when $ \theta_{i,n} $ slightly deviates away from $ \psi_{n} $. The trust region that encloses the optimal value of $ \theta_{i,n} $ is given by \cite{ref10}
	\begin{equation}\label{Eq31}
		\begin{aligned}
			& \theta_{i,n}^{*}  \in  \\
			& \begin{cases}{\left[\psi_{i,n}, \psi_{i,n}\!+\!(-1)^\lambda \Delta\right]} & \text {if } \frac{\beta_{i,n}\left(\psi_{i,n}\right)+\beta_{i,n}\left(\psi_{i,n}\!+\!\Delta\right)}{2}\!<\!\left|v_{i,n}\right|, \\
				{\left[\psi_{i,n}, \psi_{i,n}\!-\!(-1)^\lambda \Delta\right]} & \text {if } \frac{\beta_{i,n}\left(\psi_{i,n}\right)+\beta_{i,n}\left(\psi_{i,n}\!-\!\Delta\right)}{2}\!>\!\left|v_{i,n}\right|,\end{cases}
		\end{aligned}
	\end{equation}
	where $ \Delta \geq 0 $, and $ \lambda=0 $ when $ \psi_{i,n} \geq 0 $ and $ \lambda=1 $ otherwise.
	
	An accurate approximate solution can be obtained via the  one-dimensional search over the trust region, but it is computationally unaffordable. To address this issue, a closed-form approximate solution to $ \left( \ref{OptK} \right)  $ can be similarly obtained by fitting a quadratic function through three points over the trust region, i.e., $\theta_A=\varphi_n, \theta_B=\frac{2\varphi_n \pm (-1)^{\lambda} \Delta}{2}$, and $\theta_C=\varphi_n \pm (-1)^{\lambda} \Delta $. Let $f_1=f\left(\theta_A\right), f_2=f\left(\theta_B\right)$, and $f_3=f\left(\theta_C\right)$. Then, we have
	\begin{equation}\label{Eq28}
		\hat{\theta}_n^*=\frac{\theta_A\left(f_1-4 f_2+3 f_3\right)+\theta_C\left(3 f_1-4 f_2+f_3\right)}{4\left(f_1-2 f_2+f_3\right)}.
	\end{equation}

  Up to now, we finish the solution of $\mathcal{P} 3$ by the proposed AO.

	\subsection{Solution of $ \mathcal{P} 4$}
	In this subsection, we aim to solve $\mathcal{P} 4$ based on the obtained $ \mathbf{Q} $ and $\mathbf{v}^{E}$ at the above subsection.
	
	To deal with the non-convex sum-logarithms in $ \left( \ref{OptL1} \right)  $, we propose a method to decouple the logarithms based on the Lagrangian dual reformulation (LDR) technique \cite{ref46}. Following this, $ \mathcal{P}4 $ can be transformed to its equivalent problem, by introducing an auxiliary variable $\boldsymbol{\rho}=\left[\rho_{1}, \rho_{2}, \cdots, \rho_{ K} \right]^{T}$ as follows	 
	\begin{subequations}\label{OptN} 
		\begin{align}
			\mathcal{P} 4 a:   
			&\underset{\mathbf{U}, \mathbf{v}^{I}, \mathbf{P}, \mathbf{f}, \boldsymbol{\rho} }{\operatorname{max}}  g\left(\mathbf{U}, \mathbf{v}^{I},  \mathbf{P}, \mathbf{f}, \boldsymbol{\rho} \right)  \\
			 \text { s.t. }  & \left( \ref{OptA2}\right)-\left( \ref{OptA4}\right), \left( \ref{OptA7}\right), \left( \ref{OptA9}\right), 
		\end{align}
	\end{subequations}
	where $ g\left( \mathbf{U}, \mathbf{v}^{I},  \mathbf{P}, \mathbf{f}, \boldsymbol{\rho} \right) $ is given by
	\begin{equation}
		\begin{aligned}
			&g(\mathbf{U}, \mathbf{v}^{I}, \mathbf{P}, \mathbf{f}, \boldsymbol{\rho})=
			\sum_{k=1}^{K} \omega \ln \left(1\!+\!\rho_{k}\right)\!-\!\sum_{k=1}^{K}\omega\rho_{k}\\ 
			&+\sum_{k=1}^{K}  \left(1\!+\!\rho_{k}\right)\omega g_{k}(\mathbf{U}, \mathbf{v}^{I}, \mathbf{P})+\sum_{k=1}^{K}\frac{f_{k}}{C_{k}} ,
		\end{aligned}
	\end{equation}
	$ g_{k}(\mathbf{U}, \mathbf{v}^{I}, \mathbf{P}) $ is defined as
	 \begin{flalign}
			&\ g_{k}(\mathbf{U}, \mathbf{v}^{I}, \mathbf{P})= &\\
			&\ P_{k}\mathbf{h}_{k}^{H} \mathbf{u}_{k} \left(\sum_{j=1}^{K} P_{j}\mathbf{u}_{k}^{H} \mathbf{h}_{j}\left(\mathbf{u}_{k}^{H} \mathbf{h}_{j}\right)^{H}\!+\!\boldsymbol{\Xi}_{k, p}\right)^{-1}\!\!\!\!\mathbf{u}_{k}^{H} \mathbf{h}_{k}. \notag &
    \end{flalign}
     
Similar to  $ \mathcal{P} 3a^{\prime}-1 $, by introducing equality constraint $ \left( \ref{OptA3}\right) $ as a penalty term to the OF, problem $ \mathcal{P} 4 a  $ can be converted to the corresponding augmented Lagrangian (AL) problem, which is given by
  \begin{subequations}\label{OptNN} 
	\begin{align}
		\mathcal{P} 4 a^{\prime}:   
		&\underset{\mathbf{U}, \mathbf{v}^{I}, \mathbf{P}, \mathbf{f}, \boldsymbol{\rho} }{\operatorname{max}}  g\left(\mathbf{U}, \mathbf{v}^{I},  \mathbf{P}, \mathbf{f}, \boldsymbol{\rho} \right) \!-\!\iota_{2} \left\| \mathbf{v}^{I}-\mathbf{a}^{I} \right\|^{2} \\
		\text { s.t. }  & \left( \ref{OptA2}\right),\left( \ref{OptA4}\right), \left( \ref{OptA7}\right), \left( \ref{OptA9}\right), 
	\end{align}
  \end{subequations}
where $ \mathbf{a}^{I}=\left[ \beta_{1,1}\left(\theta_{1,1}\right) e^{\jmath \theta_{1,1}},..., \beta_{I,N}\left(\theta_{I,N}\right) e^{\jmath \theta_{I,N}}  \right]^{T}  $. We propose a two-layer iterative algorithm to solve $ \mathcal{P} 4 a^{\prime} $, where the inner layer solves the penalized optimization problem $ \mathcal{P} 4 a^{\prime} $ by the BCD method while the outer layer updates $ \iota_{2} $ by a factor of $\varrho>1$, until the convergence is achieved. Next, the AO scheme is still applied to solve $ \mathcal{P} 4 a^{\prime} $.

    \subsubsection{Optimizing $ \boldsymbol{\rho} $ with fixed $\left( \mathbf{U}, \mathbf{v}^{I}, \mathbf{P}, \mathbf{f} \right) $ } Given a feasible solution of $ \left(  \mathbf{U}^{\star}, \mathbf{v}^{I\star},  \mathbf{P}^{\star}, \mathbf{f}^{\star} \right)  $, the optimal $ \boldsymbol{\rho} $ can be obtained by setting  $ \partial g / \partial \rho_{k}=0 \text { for } \forall k \in \mathcal{K} $ and given by 
	\begin{equation}\label{Eq34}
		\rho_{k}^{\mathrm{opt}}=\gamma_{k}^{\star}, \quad \forall k \in \mathcal{K}.
	\end{equation}
	Note that once $ \boldsymbol{\rho} $ is decided, the objective value in $ \left( \ref{OptNN} \right)  $ is only related to the last two terms of $ g $ and the penalty term, i.e., $ \sum_{k=1}^{K}  \left(1\!+\!\rho_{k}\right)\omega g_{k}(\mathbf{U}, \mathbf{v}^{I}, \mathbf{P})+\sum_{k=1}^{K}\frac{f_{k}}{C_{k}}\!-\!\iota_{2} \left\| \mathbf{v}^{I}-\mathbf{a}^{I} \right\|^{2} $. For the MUD vector and passive beamforming design, the objective value is only related to the third term of $ g $ and the penalty term, i.e., $ \sum_{k=1}^{K}  \left(1\!+\!\rho_{k}\right) g_{k}(\mathbf{U}, \mathbf{v}^{I}, \mathbf{P})\!-\!\iota_{2} \left\| \mathbf{v}^{I}-\mathbf{a}^{I} \right\|^{2} $.
	\subsubsection{Optimizing $ \mathbf{U} $ with fixed $ \left( \mathbf{v}^{I}, \mathbf{P}, \mathbf{f}, \boldsymbol{\rho}  \right) $ } Given fixed $ \left( \mathbf{v}^{I\star}, \mathbf{P}^{\star}, \mathbf{f}^{\star}, \boldsymbol{\rho}^{\star} \right) $, the subproblem of MUD vector design at HAPs can be rewritten as 
	\begin{subequations}\label{OptO}
		\begin{align}
			&\mathcal{P} 4 a^{\prime}-1: \notag \\
			& \max _{\mathbf{U}} \overline{g}_{1}(\mathbf{U})=\sum_{k=1}^{K}  \mu_{k} g_{k}\left(\mathbf{U}, \mathbf{v}^{I\star}, \mathbf{P}^{\star} \right) \label{OptO1}\\
			& \text { s.t. }  \left\| \mathbf{u}_{b,k} \right\|^{2} \leq 1, \forall b \in \mathcal{B}, \forall k \in \mathcal{K},    \label{OptO2}
		\end{align}
	\end{subequations}
	where $\mu_{k}=\left(1+\rho_{k}^{\star}\right)\times\omega$. Since the OF in $ \left( \ref{OptO} \right)  $ is a high-dimensional sum-of-fractions, the non-convexity of $ g_{k} $ cannot be relaxed by common FP methods. To work around this non-convex high-dimensional "fractions" problem, we can adopt a method called \emph{multidimensional complex quadratic transform (MCQT)} \cite{ref46}. By introducing auxiliary variables $ \boldsymbol{\xi}=\left[\xi_{1}, \xi_{2},..., \xi_{K}  \right]  $, $ \mathcal{P} 4 a^{\prime}-1  $ can be rewritten as
	\begin{subequations}\label{OptP}
		\begin{align}
			& \mathcal{P} 4 a^{\prime}-1^{\prime}: \max _{\mathbf{U}, \boldsymbol{\xi}} \overline{g}_{2}(\mathbf{U}, \boldsymbol{\xi}) \label{OptP1} \\
			& \text { s.t. }  \left\| \mathbf{u}_{b,k} \right\|^{2} \leq 1, \forall b \in \mathcal{B}, \forall k \in \mathcal{K}, \label{OptP2}
		\end{align}
	\end{subequations}
	where 
	\begin{equation}\label{Eq37}
		\begin{aligned}
			&\overline{g}_{2}(\mathbf{U}, \boldsymbol{\xi})\!=\!\sum_{k=1}^{K} 2 \sqrt{\mu_{k} P_{k}}  \Re\left\{\boldsymbol{\xi}_{k}^{H} \mathbf{u}_{k}^{H} \mathbf{h}_{k} \right\} \\
			&\quad \quad  -\!\sum_{k=1}^{K} \boldsymbol{\xi}_{k}^{H}\left(\sum_{j=1}^{K} P_{j} \mathbf{u}_{k}^{H} \mathbf{h}_{j} \left(\mathbf{u}_{k}^{H} \mathbf{h}_{j} \right)^{H}\!+\!\boldsymbol{\Xi}_{k}\right) \boldsymbol{\xi}_{k}.
		\end{aligned}
	\end{equation}
    $ \mathbf{U} $ and $ \boldsymbol{\xi} $ can be updated alternately. $ \mathcal{P} 4 a^{\prime}-1^{\prime}  $ can be further divided into the following two independent subproblems.
	\paragraph{Fix $  \mathbf{U}  $ and solve $ \boldsymbol{\xi} $}
	By setting $ \partial \overline{g}_2 / \partial \xi_{k}=0 $, for $ \forall k \in \mathcal{K} $, the optimal $ \boldsymbol{\xi} $ is given by
		\begin{flalign}\label{Eq38}
			\xi_{k}^{\mathrm{opt}}\!=\!\sqrt{\mu_{k}P_{k}} &\left(\sum_{j=1}^{K} P_{j}\mathbf{u}_{k}^{H} \mathbf{h}_{ j}\left(\mathbf{u}_{k}^{H} \mathbf{h}_{ j}\right)^{H}\!+\!\mathbf{\Xi}_{k}\right)^{-1}\!\!\!\! \mathbf{u}_{k}^{H} \mathbf{h}_{k},\notag \\ 
			& \quad \quad \quad \quad \quad \quad \quad \quad \quad \quad \quad  \forall k \in \mathcal{K}.
		\end{flalign}
	\paragraph{Fix $ \boldsymbol{\xi} $ and solve $ \mathbf{U} $} For convenience, we first define
	\begin{equation}
		\begin{aligned}
			&\mathbf{a}_{k}\!=\!\boldsymbol{\xi}_{k}^{H} \left( \sum_{j=1}^{K} P_{j} \mathbf{h}_{j} \mathbf{h}_{j}^{H} \right)  \boldsymbol{\xi}_{k}, \\
			&\mathbf{A}_{k}=\operatorname{diag}(\mathbf{a}_{1},...,\mathbf{a}_{K}),  \quad \mathbf{v}_{k}=\sqrt{\mu_{k}P_{k}}\mathbf{h}_{k}^{H} \boldsymbol{\xi}_{k}, \\
			& \mathbf{V}=\left[\mathbf{v}_{1}, \mathbf{v}_{2}, ... ,\mathbf{v}_{K} \right]^{T}, \quad Y=\sum_{k=1}^{K} \boldsymbol{\xi}_{k}^{H} \boldsymbol{\Xi}_{k}\boldsymbol{\xi}_{k}.
		\end{aligned}
	\end{equation}
	Then, problem $ \mathcal{P} 4 a^{\prime}-1^{\prime} $ can be simplified as
	\begin{subequations}\label{OptQ}
		\begin{align}
			& \mathcal{P} 4 a^{\prime}-1^{\prime} E 1:	\notag \\
			&\underset{ \left\lbrace \mathbf{U} \right\rbrace }{\operatorname{max}} \quad	\overline{g}_{2}(\mathbf{U})=-\mathbf{U}^{H} \mathbf{A U}+\Re\left\{2 \mathbf{V} \mathbf{U}\right\}-Y,\\
			& \text { s.t. }  \mathbf{U}^{H} \mathbf{O}_{b,k} \mathbf{U} \leq 1, \forall k \in \mathcal{K},
		\end{align}
	\end{subequations}
	where $ \mathbf{O}_{b,k}=\mathbf{J}_{b,k} \otimes \mathbf{I}_{M} $, $ \mathbf{J} \in \mathbb{C}^{BK \times BK } $ is a matrix whose $ \left( (k-1)\times B+b \right) $-th row and $ \left( (k-1)\times B+b \right) $-th column is 1, and other elements are 0s. Since matrix $ \mathbf{A} $ and $ \mathbf{O}_{b,k} $ are all positive semidefinite, $ \mathcal{P} 4 a^{\prime}-1^{\prime} E 1 $ is a standard QCQP problem, which can be solved by existing CVX tools.	

	\subsubsection{Optimizing $ \mathbf{v}^{I} $ with fixed $ \left( \mathbf{U}, \mathbf{P}, \mathbf{f}, \boldsymbol{\rho} \right) $}
	Based on the given $ \left( \mathbf{U}^{\star}, \mathbf{P}^{\star}, \mathbf{f}^{\star}, \boldsymbol{\rho}^{\star} \right) $,  $ \mathcal{P} 4a^{\prime}  $ can be reformulated as the following one
	\begin{subequations}\label{OptS}
		\begin{align}
			&\mathcal{P} 4 a^{\prime}-2: 
			\max _{\mathbf{v}^{I}, \left\lbrace \theta_{i,n}^{I}\right\rbrace} \overline{g}_{3}(\mathbf{v}^{I}, \left\lbrace \theta_{i,n}^{I}\right\rbrace)\\
			&\text { s.t. } v_{i,n}=\beta_{i,n}\left(\theta_{i,n}\right) e^{\jmath \theta_{i,n}}, \forall i \in \mathcal{I}, \forall n \in \mathcal{N},  \label{OptR2} \\
			& \quad  -\pi \leq \theta_{i,n} \leq \pi, \forall i \in \mathcal{I}, \forall n \in \mathcal{N}, \label{OptS3} 
		\end{align}
	\end{subequations}
where
\begin{flalign}
	&\ \overline{g}_{3}(\mathbf{v}^{I}, \left\lbrace \theta_{i,n}^{I}\right\rbrace) = &\\
	&\ \quad \quad \quad \quad \sum_{k=1}^{K}  \mu_{k} g_{k}\left(\mathbf{v}^{I}, \mathbf{U}^{\star}, \mathbf{P}^{\star} \right)\!-\!\iota_{2} \left\| \mathbf{v}^{I}-\mathbf{a}^{I} \right\|^{2}. \notag &
\end{flalign} 
 Next, we define a new auxiliary function with respect to $ \mathbf{v}^{I} $ as
	\begin{equation}\label{Eq42}
		\mathbf{F}_{k,j}(\mathbf{v}^{I})=\sqrt{P_{j}}\mathbf{u}_{k}^{H}\left(\mathbf{h}^{d}_{j}\!+\! \mathbf{G} \mathbf{\Theta}^{I} \mathbf{h}^{r}_{j}\right).  
	\end{equation}
	Similar to $ \mathcal{P} 4 a^{\prime}-1 $, by introducing an auxiliary variable $\boldsymbol{\varpi}=\left[\varpi_{1}, \varpi_{2}, \cdots, \varpi_{K} \right]$, $ \mathcal{P} 4 a^{\prime}-2 $ can be equivalently rewritten as
	\begin{subequations}\label{OptT}
		\begin{align}
			&\mathcal{P} 4 a^{\prime}-2^{\prime}: \notag \\
 & \max _{\mathbf{v}^{I}, \left\lbrace \theta_{i,n}^{I} \right\rbrace, \boldsymbol{\varpi} } \!\! \!\!  \!\!\overline{g}_{4}(\mathbf{v}^{I}, \boldsymbol{\varpi})\!=\!\!\sum_{k=1}^{K}  \hat{g}_{k}(\mathbf{v}^{I}, \boldsymbol{\varpi})\!-\!\iota_{2} \left\| \mathbf{v}^{I}\!-\!\mathbf{a}^{I} \right\|^{2} \label{OptT1}\\
			&\quad \quad \text { s.t. }  v_{i,n}=\beta_{i,n}\left(\theta_{i,n}\right) e^{\jmath \theta_{i,n}}, \forall i \in \mathcal{I}, \forall n \in \mathcal{N},\label{OptT2}\\
			&\quad \quad \quad -\pi \leq \theta_{i,n} \leq \pi, \forall i \in \mathcal{I}, \forall n \in \mathcal{N}, \label{OptT3}
		\end{align}
	\end{subequations}
	where $ \hat{g}_{k} $ is denoted by
	  \begin{flalign}\label{Eq44}
		&\hat{g}_{k}(\mathbf{v}^{I}, \boldsymbol{\varpi})=2 \sqrt{\mu_{k}} \Re\left\{\varpi_{k}^{H} \mathbf{F}_{k, k}(\boldsymbol{\Theta}^{I})\right\} \\  &-\varpi_{k}^{H}\left(\sum_{j=1}^{K} \mathbf{F}_{k, j}(\boldsymbol{\Theta}^{I}) \mathbf{F}_{k,j}^{H}(\boldsymbol{\Theta}^{I})+\boldsymbol{\Xi}_{k}\right) \varpi_{k}, \forall k \in \mathcal{K}. \notag 
     \end{flalign}
	
	Similar to $ \mathcal{P} 4 a^{\prime}-1^{\prime} $, updating $ \mathbf{v}^{I} $ contains two steps, i.e., updating  $ \mathbf{v}^{I} $ and $ \boldsymbol{\varpi} $ in turn. Specifically, $ \mathcal{P} 4 a^{\prime}-2^{\prime} $  is further divided into two subproblems and respectively solved as follows.
	\paragraph{Fix $ \mathbf{v}^{I} $ and solve $ \boldsymbol{\varpi} $} Given a fixed $ \mathbf{v}^{I} $, the optimal  $ \boldsymbol{\varpi} $ can be obtained by solving $ \partial \overline{g}_4 / \partial \varpi_{k}=0 $, for $ \forall k \in \mathcal{K} $, and given by
	\begin{equation}\label{Eq45}
		\begin{aligned}
			\varpi_{k}^{\mathrm{opt}}\!\!=\!\!\sqrt{\mu_{k}}\left(\sum_{j=1}^{K}\! \mathbf{F}_{k, j}(\boldsymbol{\Theta}) \mathbf{F}_{k, j}^{H}(\boldsymbol{\Theta})\!+\!\boldsymbol{\Xi}_{k}\right)^{-1} \!\!\!\!\mathbf{F}_{k, k}(\boldsymbol{\Theta}).
		\end{aligned}
	\end{equation}
  \begin{figure}[htbp]
	\begin{minipage}[t]{0.42\textwidth}
		\centering
		\includegraphics[width=1.1\textwidth]{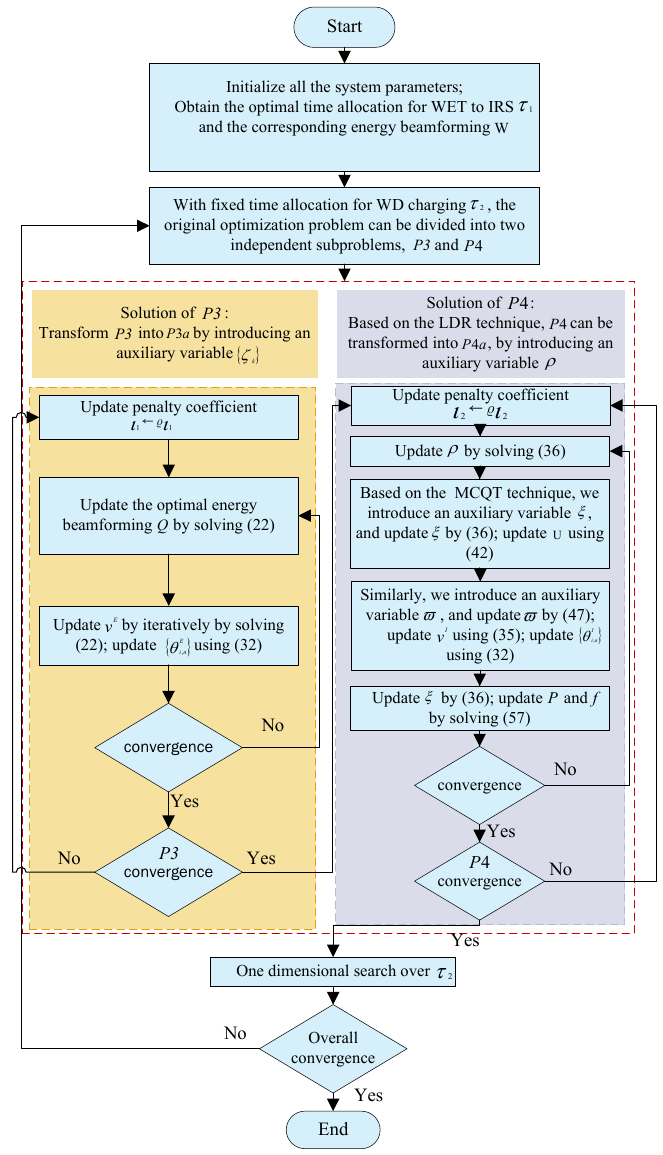}
		\caption{Block diagram of overall algorithm to solve $ \mathcal{P} 0 $.}
	\end{minipage}
\end{figure}
	\paragraph{Fix $ \boldsymbol{\varpi} $ and solve $ \mathbf{v}^{I} $} Combining with formula $ \left( \ref{Eq42} \right)  $, we simplify the expression of $ \overline{g}_{4} $ in $ \left( \ref{OptT} \right)  $ with the auxiliary function $ \mathbf{F}_{k,j} $ as follows
	\begin{equation}\label{Eq46}
		\begin{aligned}
			&\varpi_{k}^{H} \mathbf{F}_{k, j}(\mathbf{v}^{I}) \\
			&= \varpi_{k}^{H} \sqrt{P_{j}} \mathbf{u}_{k}^{H} \mathbf{h}^{d}_{j} +\varpi_{k}^{H} \sqrt{P_{j}} \mathbf{u}_{k}^{H} \mathbf{G}  \mathbf{\Theta}^{I} \mathbf{h}^{r}_{j} \\
			&= \varpi_{k}^{H} \sqrt{P_{j}} \mathbf{u}_{k}^{H} \mathbf{h}^{d}_{j}+  \varpi_{k}^{H} \sqrt{P_{j}} \mathbf{u}_{k}^{H} \mathbf{G} \operatorname{diag}\left(\mathbf{h}^{r}_{j}\right) \mathbf{v} \\
			&= c_{k, j}+\mathbf{g}_{k,j}^{H}\mathbf{v},
		\end{aligned}
	\end{equation}
	where $ c_{k, j} $ and $ \mathbf{g}_{k,j} $ are defined as
	\begin{subequations}
		\begin{align}
			&c_{k, j}=\varpi_{k}^{H} \sqrt{P_{j}} \mathbf{u}_{k}^{H} \mathbf{h}^{d}_{j},\\
			&\mathbf{g}_{k,j}=\left( \varpi_{k}^{H} \sqrt{P_{j}} \mathbf{u}_{k}^{H} \mathbf{G} \operatorname{diag}\left(\mathbf{h}^{r}_{j}\right)\right) ^{H}.
		\end{align}
	\end{subequations}
	Then, substituting $ \left( \ref{Eq46} \right)  $ into $ \left( \ref{Eq44} \right)  $, yields
	\begin{equation}
		\begin{aligned}
			&\hat{g}_{k}(\mathbf{v}^{I})=2 \sqrt{\mu_{k}} \mathfrak{\Re}\left\{c_{k, k}+ \mathbf{g}_{k, k}^{H}\mathbf{v}^{I} \right\}\\
			& -\sum_{j=1}^{K}\left(c_{k, j}+ \mathbf{g}_{k, j}^{H} \mathbf{v}^{I}\right)\left(c_{k, j}^{*}+\left( \mathbf{v}^{I}\right) ^{H}\mathbf{g}_{k, j} \right)-\varpi_{k}^{H} \boldsymbol{\Xi}_{k} \varpi_{k}.
		\end{aligned}
	\end{equation}
	Therefore, $ \overline{g}_{4} $ in $ \left( \ref{OptT1}\right)  $ can be simplified as
	\begin{equation}
		\overline{g}_{4}(\mathbf{v}^{I})\!=\!-\!\left( \mathbf{v}^{I}\right) ^{H} \boldsymbol{\Lambda} \mathbf{v}^{I}\!+\!\Re\left\{2 \boldsymbol{\nu} \mathbf{v}^{I}\right\}\!-\!\zeta \!-\!\iota_{2} \left\| \mathbf{v}^{I}\!-\!\mathbf{a}^{I} \right\|^{2}\!\!,
	\end{equation}
	where
	\begin{equation}
		\begin{aligned}
			\boldsymbol{\Lambda}=& \sum_{k=1}^{K} \sum_{j=1}^{K} \mathbf{g}_{k, j} \mathbf{g}_{k, j}^{H}, \\
			\boldsymbol{\nu}=& \sum_{k=1}^{K} \sqrt{\mu_{k}} \mathbf{g}_{k,k}^{H}-\sum_{k=1}^{K} \sum_{j=1}^{K} c_{k, j}^{*} \mathbf{g}_{k, j}^{H}, \\
			\zeta=& \sum_{k=1}^{K} \sum_{j=1}^{K}\left|c_{k, j}\right|^{2}+\sum_{k=1}^{K} \varpi_{k}^{H} \boldsymbol{\Xi}_{k} \varpi_{k}-2 \sum_{k=1}^{K} \sqrt{\mu_{k}} \mathfrak{\Re}\left\{c_{k, k}\right\}.
		\end{aligned}
	\end{equation}
	As such, the non-convex $ \mathcal{P} 4 a^{\prime}-2^{\prime} $ has been converted to a convex one, which is given by
	\begin{subequations}\label{OptV}
		\begin{align}
			&\mathcal{P} 4 a^{\prime}-2^{\prime} E 1: \notag \\
			&\quad \min _{\mathbf{v}^{I} } \left( \mathbf{v}^{I}\right) ^{H} \boldsymbol{\Lambda} \mathbf{v}^{I}\!-\!\Re\left\{2 \boldsymbol{\nu} \mathbf{v}^{I}\right\}\!+\!\iota_{2} \left\| \mathbf{v}^{I}-\mathbf{a}^{I} \right\|^{2}\\
			&\quad \text { s.t. } \left| v_{i, n}^{I} \right| \leq 1, \forall i \in \mathcal{I}, \forall n \in \mathcal{N}.
		\end{align}
	\end{subequations}
Since matrix $ \boldsymbol{\Lambda} $ is positive semidefinite, $\mathcal{P} 4 a^{\prime}-2^{\prime} E 1$ is convex and can be solved by the standard convex optimization tool. Then, based on the obtained $ \mathbf{v}^{I} $, we update $ \left\lbrace \theta_{i,n}^{I} \right\rbrace $  using $ \left( \ref{Eq28} \right)  $. 

	\begin{figure*}[b]
		{\noindent}
		\rule[-10pt]{18.07cm}{0.1em}
		\begin{equation}\label{Eq53}
			\begin{aligned}
				\overline{g}_{5}(\mathbf{P})&=\sum_{k=1}^{K} 2 \sqrt{\mu_{k} P_{k}}  \Re\left\{\boldsymbol{\xi}_{k}^{H} \mathbf{u}_{k}^{H} \mathbf{h}_{k} \right\} 
				-\sum_{k=1}^{K} \boldsymbol{\xi}_{k}^{H}\left(\sum_{j=1}^{K} P_{j} \mathbf{u}_{k}^{H} \mathbf{h}_{j} \left(\mathbf{u}_{k}^{H} \mathbf{h}_{j} \right)^{H}+\boldsymbol{\Xi}_{k}\right) \boldsymbol{\xi}_{k} \\
				&=\sum_{k=1}^{K} 2 \sqrt{\mu_{k}}  \Re\left\{\boldsymbol{\xi}_{k}^{H} \mathbf{u}_{k}^{H} \sqrt{P_{k}}\mathbf{h}_{k} \right\} 
				-\sum_{k=1}^{K} \boldsymbol{\xi}_{k}^{H} \mathbf{u}_{k}^{H} \left( \sum_{j=1}^{K} \sqrt{P_{j}} \mathbf{h}_{j} \mathbf{h}_{j}^{H} \sqrt{P_{j}}\right)\mathbf{u}_{k}\boldsymbol{\xi}_{k} -\boldsymbol{\xi}_{k}^{H}\boldsymbol{\Xi}_{k} \boldsymbol{\xi}_{k}
			\end{aligned}
		\end{equation}
	\end{figure*}	

	\subsubsection{Optimizing $ \mathbf{P} $ and $ \mathbf{f} $ with fixed $ \left(  \mathbf{U}, \mathbf{v}^{I}, \mathbf{f}, \boldsymbol{\rho} \right) $}  For given $ \left( \mathbf{U}^{\star} , \mathbf{v}^{I \star}, \mathbf{f}^{\star}, \boldsymbol{\rho}^{\star } \right) $ and $ \boldsymbol{\xi}^{\star} $, $ \overline{g}_{2} $ can be reformulated as $  \overline{g}_{5}  $ at the bottom of next page. Then, for simplification, we define
	\begin{equation}\label{Eq41} 
		\begin{aligned}
			& \mathbf{b}=\sum_{k=1}^{K} \mathbf{u}_{k}\boldsymbol{\xi}_{k}\boldsymbol{\xi}_{k}^{H} \mathbf{u}_{k}^{H}, \quad \mathbf{B}=\mathbf{I}_{K}\otimes \mathbf{b}, \\
			& \mathbf{c}_{k}=2 \sqrt{\mu_{k}} \boldsymbol{\xi}_{k}^{H} \mathbf{u}_{k}^{H}, \quad \mathbf{C}=\left[\mathbf{c}_{1}, ... ,\mathbf{c}_{K} \right], \\
			& \mathbf{P}^{'}=\left[ \left( \sqrt{P}_{1}\mathbf{h}_{1}\right) ^{T},..., \left( \sqrt{P}_{K}\mathbf{h}_{K}\right) ^{T} \right] ^{T},\\
			& \chi=\boldsymbol{\xi}_{k}^{H}\boldsymbol{\Xi}_{k} \boldsymbol{\xi}_{k}.
		\end{aligned}
	\end{equation}
	By substituting $ \left( \ref{Eq41}\right)  $ to $ \left( \ref{Eq53} \right)  $, we obtain
	\begin{equation}
		\begin{aligned}
			\overline{g}_{5}(\mathbf{P})=-\mathbf{P}^{'H} \mathbf{B}\mathbf{P}^{'}+\Re\left\{ \mathbf{C} \mathbf{P}^{'}\right\}-\chi.
		\end{aligned}
	\end{equation}
	Therefore, the subproblem of uplink transmit power and local computing frequency design at WDs can be formulated as
	\begin{subequations}\label{OptR}
		\begin{align}
			& \mathcal{P} 4 a^{\prime}-3:	\max _{ \mathbf{P}, \mathbf{f}} \notag \\
			& \omega \left( \sum_{k=1}^{K} \ln \left(1\!+\!\rho_{k}\right)\!-\!\sum_{k=1}^{K}  \rho_{k}\right)\!+\!\overline{g}_{5}(\mathbf{P})\!+\!\sum_{k=1}^{K} f_{k}/C_{k} \label{OptR1}\\
			&\text { s.t. }  0 \leq f_{k} \leq f_{k, \max }, \quad \forall k \in \mathcal{K}, \label{OptR2} \\
			&t_{1}\left[\kappa f_{k}^{2}+\left(P_{k}+P_{c}\right)\right] \leq  E_{k}\left(\tau_{2},\mathbf{Q} \right),  \forall k \in \mathcal{K}. \label{OptR3}
		\end{align}
	\end{subequations}
 Since matrix $ \mathbf{B} $ is positive semidefinite, the last term of OF is linear, and contraints $ \left( \ref{OptR2} \right)  $ and $ \left( \ref{OptR3} \right)  $ are linear and convex. Therefore, $ \mathcal{P} 4 a^{\prime}-3 $ is a convex optimization problem, which can be solved by the convex optimization tool.
 
 Now we have the complete solution for the original optimization problem $\mathcal{P} 0$, and the corresponding flow diagram is depicted in Fig. 4. 
 
 As the diagram shows, after obtaining the optimal time allocation for IRS charging and the corresponding energy beamforming, and fixing the time allocation for WD charging, the original optimization problem can be devided into two independent subproblems, i.e., $ \mathcal{P} 3 $ and $ \mathcal{P} 4 $. The first subproblem $ \mathcal{P} 3 $ aims to optimize the WD charging setting. To facilitate the solution, $ \mathcal{P} 3 $ is transformed into $ \mathcal{P} 3 a ^{'} $ by introducing auxiliary variable $ \left\lbrace \zeta_{k} \right\rbrace $. Due to the nonconvex reflecting coefficient constraint of IRS, the reformulated $ \mathcal{P} 3 a ^{'} $ is still difficult to solve directly. To proceed it, we propose a penalty-based method by adding a penalty term to the OF, and $ \iota_{1} $ is the penalty parameter. Then, $ \mathcal{P} 3 a ^{'} $ can be solved by our proposed two-layer iterative algorithm, where the inner layer updates energy beamforming $ \mathbf{Q} $ and passive beamforming $ \mathbf{v}^{E} $ while the outer layer updates $ \iota_{1} $, until the convergence is achieved. The second subproblem $ \mathcal{P} 4 $ aim to optimize the computing setting, and based on the LDR technique, the nonconvex sum-logarithms in its OF is decoupled by introducing an auxiliary variable $\boldsymbol{\rho}$. Similar to $ \mathcal{P} 3 a ^{'} $, we add the nonconvex reflecting coefficient constraint of IRS as a penalty term to the OF, and $ \iota_{2} $ is the penalty parameter. Similarly, the reformulated subproblem $ \mathcal{P} 4a^{'} $ can be solved by our proposed two-layer algorithm framework, in which the inner layer algorithm iteratively update auxiliary variable $\boldsymbol{\rho}$, MUD matrix $ \mathbf{U} $, passive beamforming $ \mathbf{v}^{I} $, uplink transmit power $ \mathbf{P} $ and local computing frequency $ \mathbf{f} $, and the outer layer algorithm update penalty coefficient $ \iota_{2} $, until the convergence is achieved. Finally, the optimal time allocation $ \tau_{2} $ is obtained by one dimensional search method. 
			
\section{Numerical Results}		
 \begin{figure}[htbp]
	\begin{minipage}[t]{0.42\textwidth}
		\centering
		\includegraphics[width=1.1\textwidth]{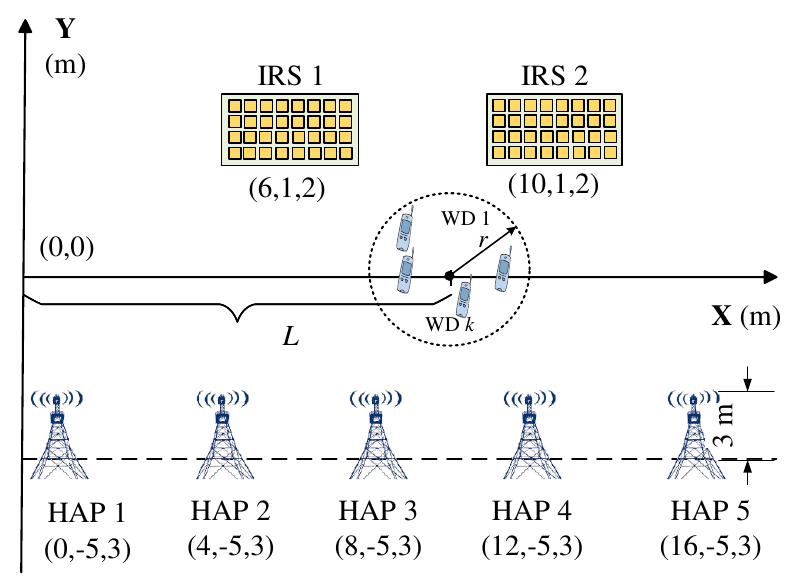}
		\caption{The simulation scenario with five HAPs and two IRSs.}
	\end{minipage}
\end{figure}
 In this section, numerical results are provided to evaluate the performance of the proposed algorithm. Similar to \cite{ref47}, we consider a three dimensional coordinate system as shown in Fig.~5, which consists of five HAPs, two IRSs, and four WDs. Here, two IRSs are deployed on the user side to improve network capacity. We assume that the $ b $-th HAP is located at $ \left(4\left( b-1\right) \mathrm{m}, -5 \mathrm{m}, 3 \mathrm{m} \right)  $, two IRSs are located at $ \left( 6 \mathrm{m}, 1 \mathrm{m}, 2 \mathrm{m} \right)  $ and $ \left( 10 \mathrm{m}, 1 \mathrm{m}, 2 \mathrm{m} \right)  $, and $ K $ WDs are uniformly and randomly distributed in a cluster, which is centered at $ (L \mathrm{m}, 0 \mathrm{m}, 1 \mathrm{m}) $ with radius 1 m. The simulation parameters are summarized in $ \text { Table II} $.

\begin{table}[htbp]
	\centering
	\footnotesize
	\renewcommand{\arraystretch}{1.5}
	\setlength{\tabcolsep}{3pt}
	\caption{Simulation Parameters.}
	\label{tab:my_label}
	\scalebox{.96}{
		\begin{tabular}{ll}
			\toprule
			\textbf{Parameters} & \textbf{Values}  \\ \midrule
			Number of WD, $ K $  & 4  \\
			Number of HAP, $ B $    &  5  \\
			Number of HAP's antennas, $M$ &  2  \\
			Number of IRS, $ I $        &  2  \\ 
			Number of IRS reflection element, $ N $  & 10  \\
			Maximum transmit power of HAP          &  100W \\
			Time block length, $ T $   &  1s \\
			System bandwidth, $ \omega $ & 1MHz \\      
			\tabincell{l}{Local computing capability of the  $ k $-th WD, $ f_{k, \max } $}  &   $ 1 \times 10^{8}  $ cycles/s  \\                           
			Computational complexity, $ C_{k} $ &  500 cycles/bit \\
			Energy efficiency coefficient, $ \kappa $ &  $ 1 \times 10^{-28} $ \\
			Energy conversion efficiency, $ \eta  $ & 0.8 \\
			Noise spectral density, $ N_{0} $ & 40 dBm\\
			Path loss exponent, $ \kappa_{HU}/\kappa_{HI}/\kappa_{IU} $ & 3.5/2.2/2.8 \\
			\tabincell{r}{Power consumption of each reflection element, $ \mu $}  & 1mW \\
			\bottomrule
		\end{tabular}
	}
\end{table}
 As for the communications channel, both the small scale fading and large scale path loss are considered. We define $ d_{HU} $, $ d_{HI} $, and $ d_{IU} $ as the distance between HAP and WD, HAP and IRS, IRS and WD. Thus, the distance-dependent large scale path loss model is given by  
\begin{equation}\label{Eq55}
	L\left( d \right)=C_{0}\left( \frac{d}{d_{0}} \right) ^{-\kappa}\!, \quad	d \in \left\lbrace d_{HU}, d_{HI}, d_{IU}   \right\rbrace, 
\end{equation}	
where $ C_{0}=-30dB $ denotes the path loss at a reference distance $ d_{0}=1 $ m. $ d $ refers to the individual channel distance, and $ \kappa $ is the path loss exponent. Here we set the path loss exponent of the HAP-WD link, HAP-IRS link, and IRS-WD link to $ \kappa_{HU}=3.5 $, $ \kappa_{HI}=2.2 $, and $ \kappa_{IU}=2.8 $ \cite{ref47}, respectively. For the small scale fading, we consider a Rician fading channel model for all involved channels, and the channel $ \mathbf{H} $ is modeled as 
\begin{equation}
	\mathbf{H}=\sqrt{\frac{\beta_{\mathrm{HU}}}{1+\beta_{\mathrm{HU}}}} \mathbf{H}^{\mathrm{LoS}}+\sqrt{\frac{1}{1+\beta_{\mathrm{HU}}}} \mathbf{H}^{\mathrm{NLoS}},
\end{equation}
where $ \beta_{\mathrm{HU}} $ stands for the Rician factor, $ \mathbf{H}^{\mathrm{LoS}} $ and $ \mathbf{H}^{\mathrm{NLoS}} $ refer to the LoS deterministic component and the non-LoS Rayleigh fading component, respectively. $ \mathbf{H} $ is reduced to a Rayleigh fading channel model when $ \beta_{\mathrm{HU}}=0 $, a LoS channel model when $ \beta_{\mathrm{HU}} \rightarrow \infty $. Then, for the HAP-WD channel, we need to multiply  $ \mathbf{H} $ by  the distance-dependent large scale path loss $ L\left( d \right) $ in $ \left( \ref{Eq55}\right)  $. Similarly, the HAP-IRS and IRS-WD channels can also be generated by the above procedure with $  \beta_{\mathrm{HI}} $ and $ \beta_{\mathrm{IU}} $ denoting the Rician factors of them. Here we assume $  \beta_{\mathrm{HI}} \rightarrow \infty $, $ \beta_{\mathrm{IU}} \rightarrow \infty $, and $ \beta_{\mathrm{HU}}=0 $. Note that the energy efficiency coefficient $ \kappa $ setting follows \cite{ref7, ref26}, the local computing capability of the $ k $-th WD $ f_{k,max} $ setting follows \cite{ref29}, and the setting of other parameters in Table II is given by \cite{ref7} or \cite{ref47}.

For comparison, the following four baseline schemes are considered.
\begin{itemize}
	\item \emph{Upper bound:} solve $ \mathcal{P} 0 $ with $ \beta_{min}=1 $ using the proposed method.
	\item \emph{Ideal IRS assumption:} the phase shift design under ideal IRS assumption is applied to the practical IRS model.
	\item \emph{Full offloading:} all computation tasks of each WD are offloaded to the HAP and performed on the edge server.
	\item \emph{No IRS:} the conventional communication network without IRS. 
\end{itemize}

\begin{figure*}[htbp]
	\quad \quad \subfigure[]{
		\begin{minipage}[t]{0.4\textwidth}
			\centering
			\includegraphics[width=1.1\textwidth]{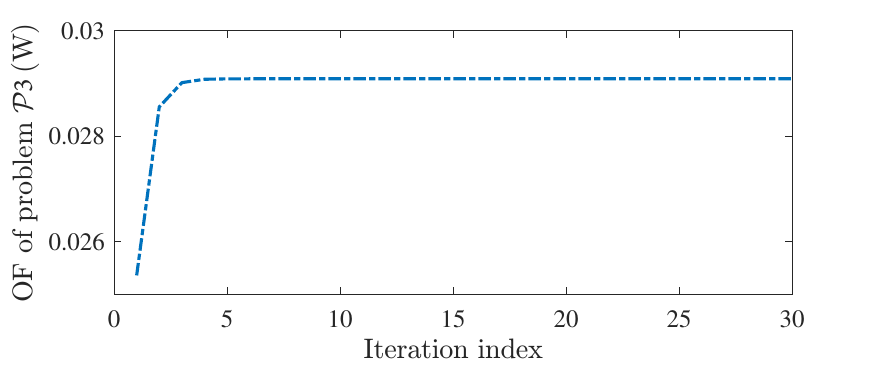}
		\end{minipage}
		}
\quad 
	\subfigure[]{
		\begin{minipage}[t]{0.4\textwidth}
			\centering
			\includegraphics[width=1.1\textwidth]{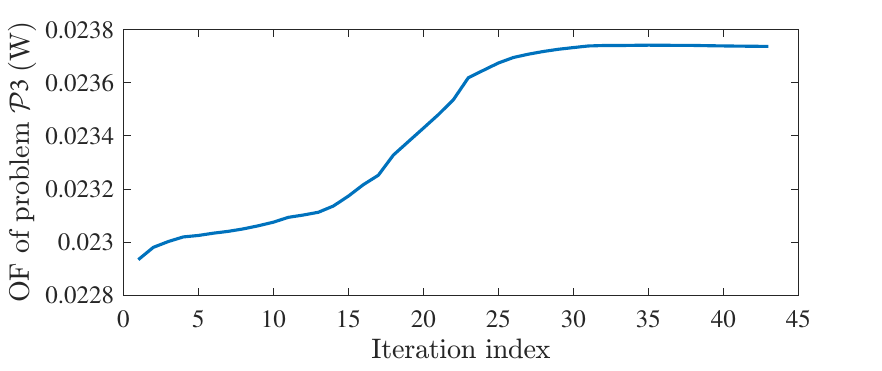}
		\end{minipage}
	}

   \quad \quad \subfigure[]{
    	\begin{minipage}[t]{0.4\textwidth}
    		\centering
    		\includegraphics[width=1.1\textwidth]{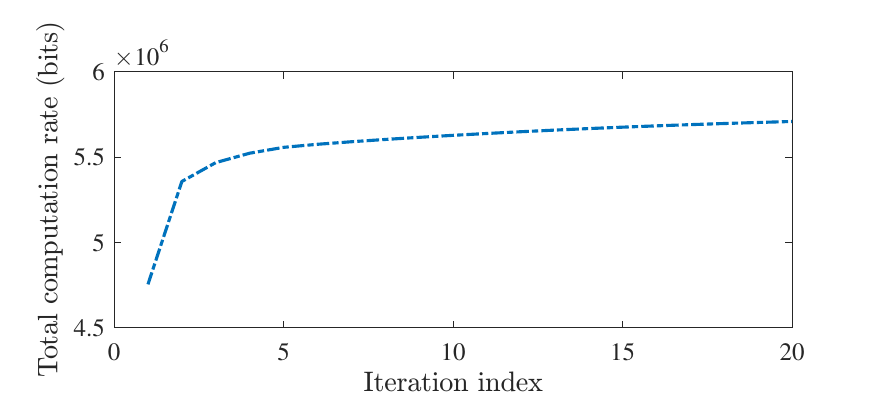}
    	\end{minipage}
    }
\quad 
   \subfigure[]{
   	\begin{minipage}[t]{0.4\textwidth}
   		\centering
   		\includegraphics[width=1.1\textwidth]{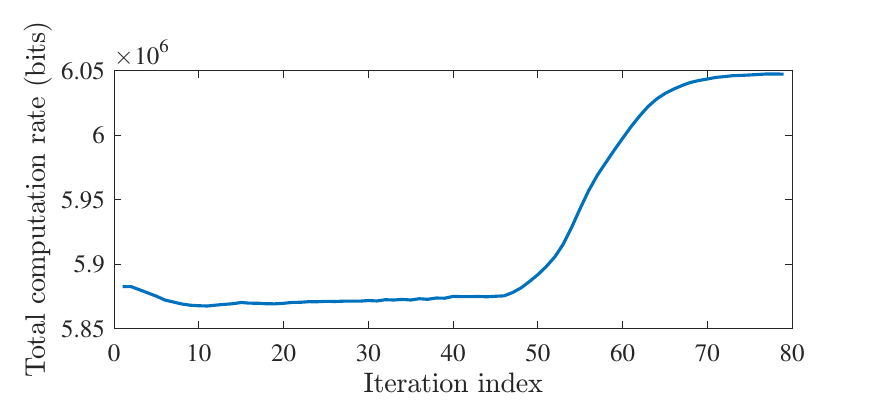}
   	\end{minipage}
   }
  \caption{Convergence behavior of the proposed algorithms. (a) Inner loop for solving $\mathcal{P} 3$; (b) Outer loop for solving $\mathcal{P} 3$; (c) Inner loop for solving $\mathcal{P} 4$; (d) Outer loop for solving $\mathcal{P} 4$; }
  \end{figure*}

\begin{figure*}[htbp]
	\begin{minipage}[t]{0.42\textwidth}
		\centering
		\includegraphics[width=1.2\textwidth]{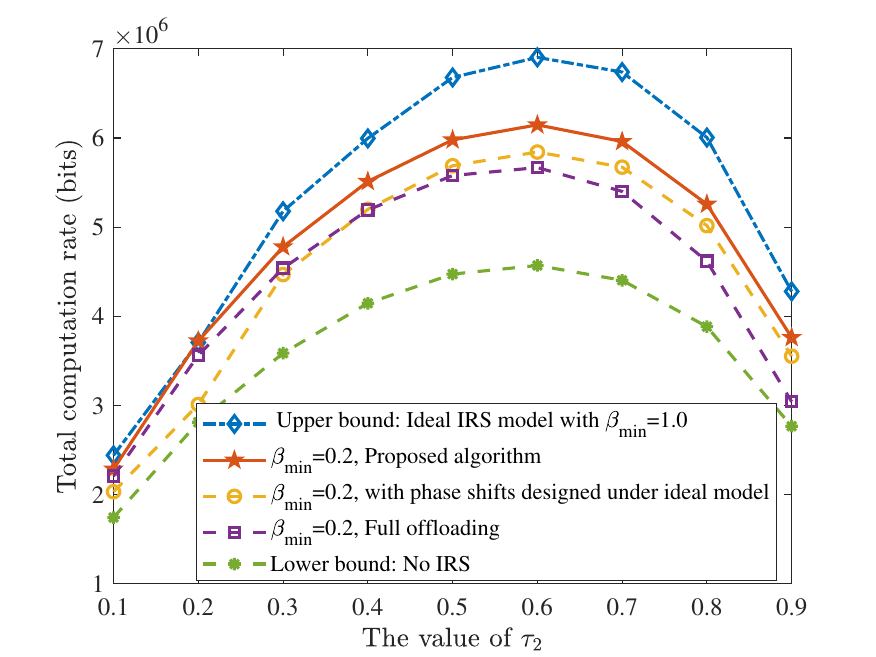}
		\caption{ Total computation rate versus $ \tau_{2}$.}
	\end{minipage}
    \qquad \qquad
	\begin{minipage}[t]{0.42\textwidth}
	   \centering
   	   \includegraphics[width=1.2\textwidth]{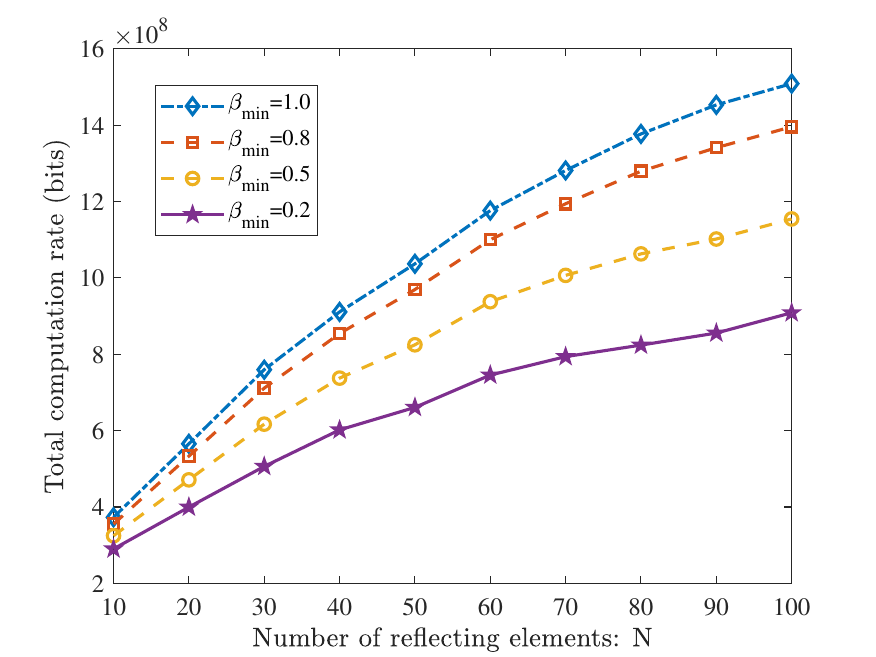}
	   \caption{Total computation rate versus the number of IRS elements.}
    \end{minipage}
\end{figure*}	

\begin{figure*}[htbp]
	\begin{minipage}[t]{0.42\textwidth}
		\centering
		\includegraphics[width=1.2\textwidth]{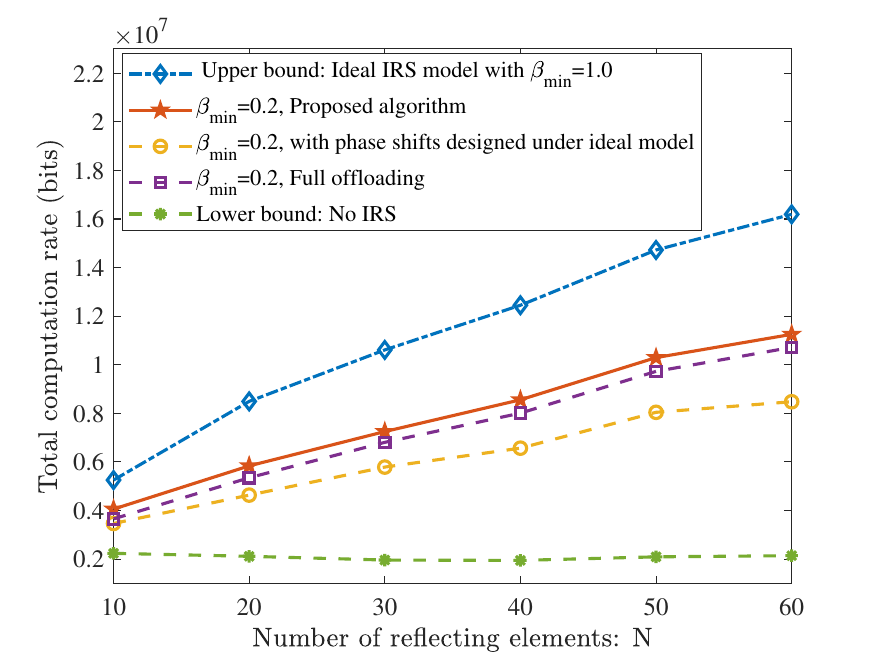}
		\caption{Total computation rate versus the number of IRS elements.}
	\end{minipage}
	\qquad \qquad
	\begin{minipage}[t]{0.42\textwidth}
		\centering
		\includegraphics[width=1.2\textwidth]{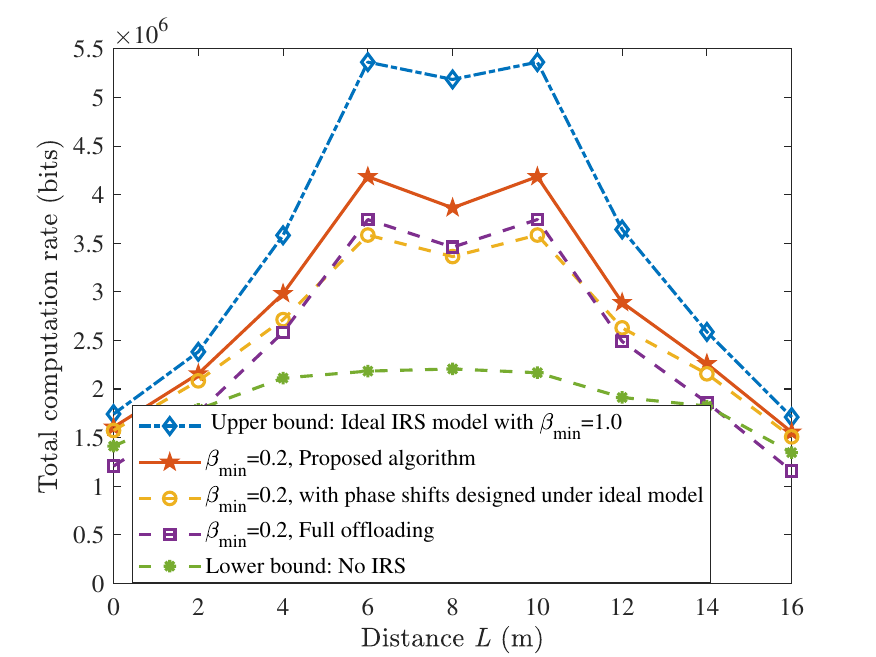}
		\caption{Total computation rate versus the distance $ L $.}
	\end{minipage}
\end{figure*}	

\begin{figure*}[htbp]
	\begin{minipage}[t]{0.42\textwidth}
		\centering
		\includegraphics[width=1.2\textwidth]{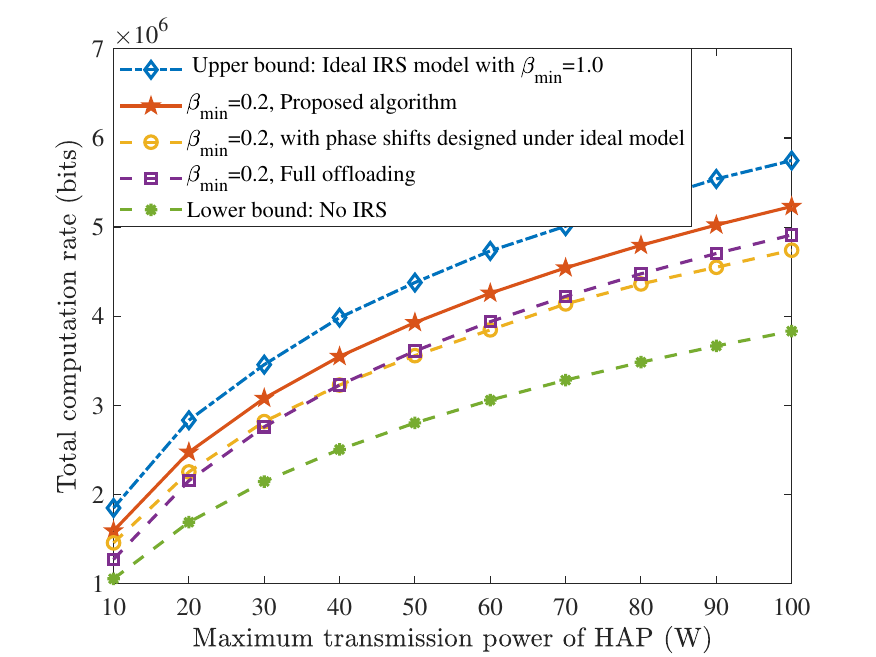}
		\caption{Total computation rate versus maximum transmit power of HAP.}
	\end{minipage}
	\qquad \qquad
	\begin{minipage}[t]{0.42\textwidth}
		\centering
		\includegraphics[width=1.2\textwidth]{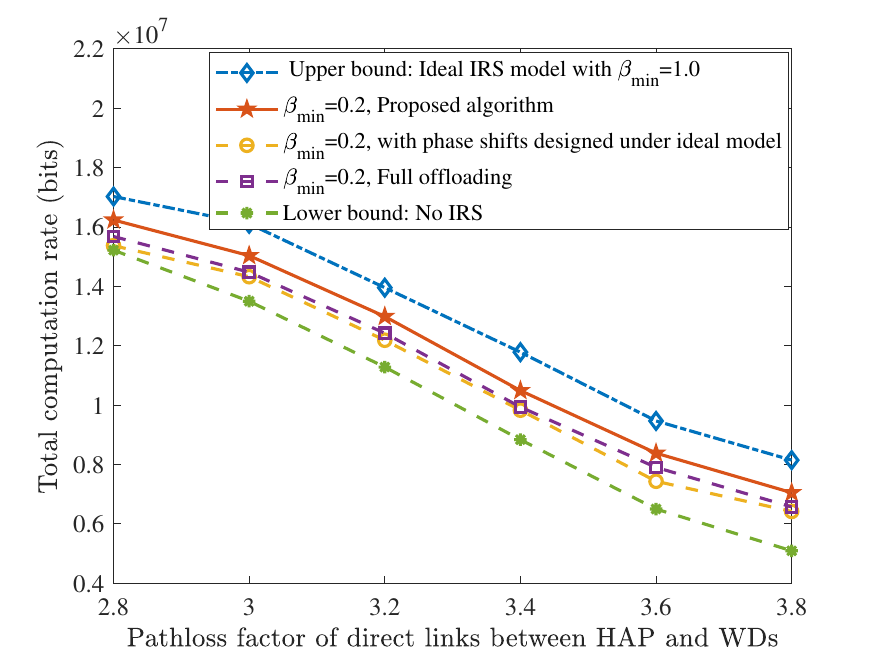}
		\caption{Total computation rate versus path loss exponent $ \kappa_{HU} $.}
	\end{minipage}
\end{figure*}

\emph{1) Convergence Behavior:} Figs. 6 (a)-(b) show the convergence behavior for solving $\mathcal{P} 3 $. It is observed that the proposed algorithm for downlink energy beamforming and IRS passive beamforming design converges within 5 iterations, while the outer loops that is related to the initial value of $ \iota_{1} $, converges within 30 iterations. Figs. 6 (c)-(d) present the convergence behavior for solving $\mathcal{P} 4$. Similarly, it can be seen that the inner loop for optimizing MUD vector, IRS passive beamforming, offloading power, and local computing frequency can reach convergence within 10 iterations, while the outer loop converges within 70 iterations. 

\emph{2) Selection of the Time Allocation:} Combining $ \left( \ref{OptA1} \right)  $ and $ \left( \ref{Eq16} \right)  $, the time allocation satisfies $ \tau_{2}+t_{1}=T-\tau_{1}^{\star} $. In order to find an appropriate time allocation, we depict the total computation rate versus the time allocation $ \tau_{2} $ in Fig. 7. It is observed that $ \mathcal{P} 2 $ is convex with respect to $ \tau_{2} $, and this means that, there exists a $ \tau_{2}^{\star} $ that maximizes the total computation bits. Here, we use the one-dimensional search method to find the optimal $ \tau_{2} $.

\emph{3) Impact of the Parameter $ \beta_{\min} $:} In Fig. 8, we show the total computation rate against the number of IRS elements $ N $. Specifically, the IRS phase shift design for ideal IRS is applied to the practical IRS with $ \beta_{\min}=0.8 $, $ \beta_{\min}=0.5 $, and $ \beta_{\min}=0.2 $. One can observe that as the number of IRS element $ N $ increases, the performance gap between ideal case $ \left( \beta_{\min}=1 \right)  $ and practical cases $\left(\beta_{\min }=0.8, \beta_{\min }=0.5\right.$, and $\left.\beta_{\min }=0.2\right)$ first increases and then approaches a constant, which is determined by $ \beta_{\min} $ and $ \alpha $ \cite{ref10}. This is because that when $ N $ is moderate, the signal power received from the HAP-WD link is comparable to that of the IRS-WD link, thus the performance loss due to the imperfect hardware of IRS is more pronounced as $ N $ increases. Whereas, when $ N $ is sufficiently large, the reflection link dominates. Hence the performance loss arising from the IRS reflection imperfection converges to a constant. This observation implies that the promising squared power scaling order, i.e., $\mathcal{O}\left(N^2\right)$ unveiled in \cite{ref48} under the ideal phase shift model, still holds for practical case. Furthermore, this indicates that it is crucial to take into consideration the practical IRS phase shift model for the passive beamforming design in practical IRS-assisted systems.

\emph{4) Impact of the Number of IRS Elements $ N $:} Fig. 9 presents the total computation rate versus the number of reflecting elements for all considered schemes. It is observed that the performance gap between ideal phase shift model and practical phase shift model increases as the number of reflecting elements increases. This is due to the fact that, with $ N $ increases, the IRS-reflected signal power dominates in the total received power of WDs, thus, the performance loss due to the IRS hardware imperfection is more pronounced. Noted that, too more IRS elements will make the channel estimations challenging. Therefore, it is essential to choose an appropriate IRS element number.

\emph{5) Impact of the Distance $ L $:} From Fig. 10, we can observe that for the schemes with IRS deployed, there are two obvious peaks at $ L=6 $ m and $ L=8 $ m. This is because when WDs approach one of the two IRSs, WDs can receive strong signals reflected from IRSs. While for the conventional scheme without IRS, the two peaks will not appear. When WDs approach one of the two IRSs, the performance gap between the ideal IRS case $ \left( \beta_{\min}=1 \right)  $ and practical IRS case $ \left( \beta_{\min}=0.2 \right) $ increases. This is because the received signals power reflected from IRS play a dominant role, while when WDs are far from IRSs, the contribution of IRS becomes limited.

\emph{6) Impact of HAP's Maximum Transmit power $ P_{\max} $:} By fixing $ L=6 $ m, we plot the total computation rate against the maximum transmit power of HAP in Fig. 11. We can observe that the total computation rate for all considered schemes increases monotonically with the increase of HAP's maximum transmit power. Obviously, this is due to the fact that WDs will harvest more energy when HAPs adopt a higher transmit power, which will in turn enable WDs to perform more computation data via local computing and task offloading. Furthermore, we see that, compared to the existing schemes, the proposed algorithm can achieve a better performance. Particularly, the performance gain achieved by the proposed algorithm over the conventional scheme without IRS increases with the increase of $ P_{\max} $. This indicates that when WDs can harvest more energy with a larger $ P_{\max} $, the reflection link gain by employing IRS can improve the task offloading rate significantly.

\emph{7) Impact of Path Loss exponent:} In Fig. 12, we present the total computation rate against the path loss exponent of direct links between WDs and HAPs. We see that the total computation rate decreases with the increase of path loss exponent. This is because that a higher path loss exponent will result in a lower channel gain for the composite channel between HAPs and WDs. Besides, compared to the full offloading scheme and conventional scheme without IRS, the proposed IRS-assisted wireless powered MEC system can achieve 7 $ \% $ and 38$ \% $ performance gain, respectively.

\emph{8) Impact of the Number of WDs:} In Fig. 13, we plot the total computation rate versus the number of WDs. We see that although the total computation rate increases from $ K=2 $ to $ K=8 $, the increasing slope of the schemes with IRS is larger than the conventional scheme without IRS. This observation corroborates that employing IRS in the wireless communication system can achieve a higher throughput by carefully designing the IRS passive beamforming.

\emph{9) Impact of CSI Error Parameter:} Generally, due to the high-dimensional channels introduced by IRS, the channel estimation is very challenging for the IRS-assisted  communication system. We analyze the robustness of the proposed algorithm to CSI error, and model the imperfect channels as 
\begin{equation}
	\hat{h}=h+e,
\end{equation}
where $ \hat{h} $ and $ h $ stand for the practically estimated channel and the real channel, respectively. $ e $ denotes the estimation error with Gaussian distribution, which satisfy zero mean and variance $ \sigma_e^2 $, i.e., $e \sim \mathcal{C} N\left(0, \sigma_e^2\right)$. We define $ \delta $ as the ratio of error power $ \sigma_e^2 $ to the channel gain $ |h|^2 $, i.e., $ \delta \triangleq \sigma_e^2/|h|^2 $, to characterize the level of CSI error. In Fig. 14, we depict the total computation rate against the CSI error parameter $ \delta $. We see that the performance loss increases with  $ \delta $. In particular, for the proposed algorithm, compared to the perfect CSI without error ($ \delta=0 $), the system performance suffers a loss of 6 $ \% $ when $ \delta=0.1 $, and a loss of 21 $ \% $ when $ \delta=0.3 $. Thereby, the proposed algorithm shows strong robustness to the CSI error.

\section{Conclusion}
Based on the practical IRS phase shift model, this paper studied the resource management problem in an IRS-assisted wireless powered MEC network. The downlink/uplink passive beamforming, downlink energy beamforming and uplink MUD vector at HAP, task offloading power and local computing frequency at WDs, and time allocation for WET and computing were jointly optimized to maximize the total computation rate under energy casuality constraints of IRSs and WDs. A sophisticated algorithm was provided to optimize those parameters both in the WET and computing phases.  Finally, numerical results showed that as system parameters change (such as the number of IRS elements, location, and HAP transmit power), the performance loss caused by the imperfect hardware of IRS is more pronounced. Therefore, it is crucial to take into consideration the practical IRS phase shift model for the passive beamforming design in practical IRS-assisted systems. Furthermore, simulation results demonstrated that, compared to the conventional method with ideal phase shift model, which is widely used in the literature, our proposed algorithm can achieve a significant performance gain.

	 \begin{figure*}[htbp]
			\begin{minipage}[t]{0.42\textwidth}
				\centering
				\includegraphics[width=1.2\textwidth]{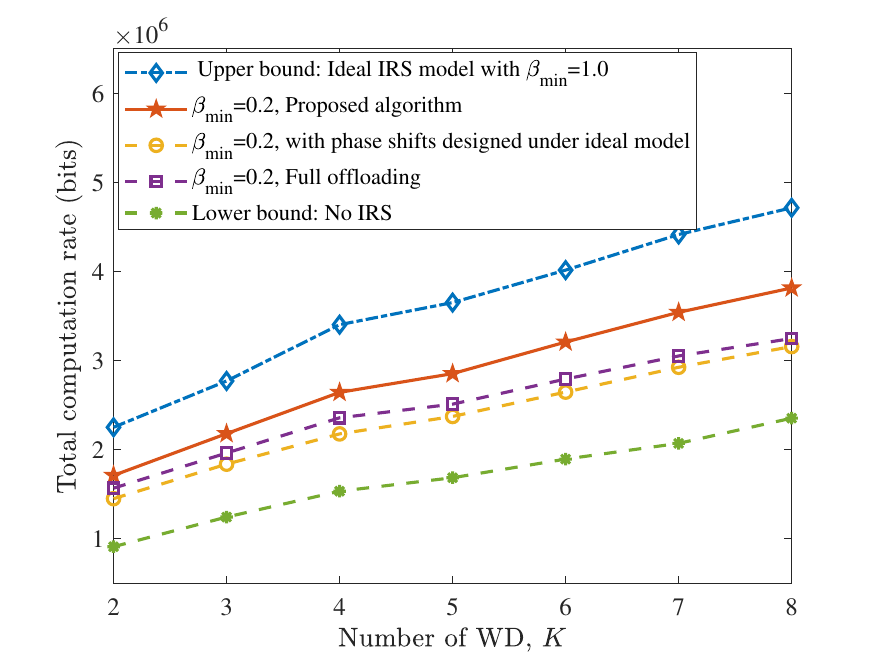}
				\caption{Total computation rate versus the number of WDs $ K $.}
			\end{minipage}
			\qquad \qquad
			\begin{minipage}[t]{0.42\textwidth}
				\centering
				\includegraphics[width=1.2\textwidth]{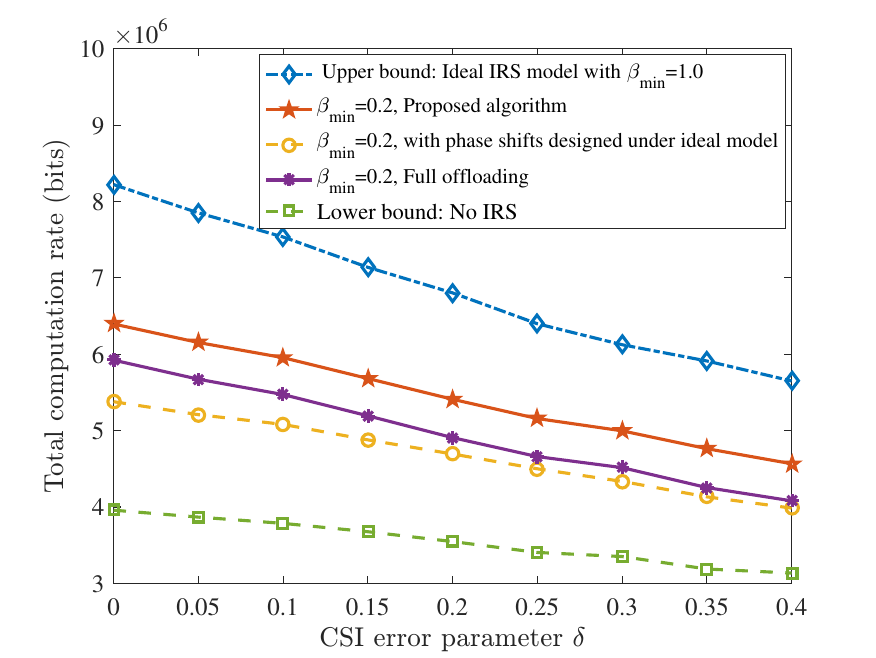}
				\caption{Total computation rate versus CSI error parameter $ \delta $.}
			\end{minipage}
	 \end{figure*}


\begin{thebibliography}{99}
		\bibliographystyle{IEEEtran}
		
		\bibitem{ref1}
		L. Chettri and R. Bera, “A comprehensive survey on Internet of Things (IoT) toward 5G wireless systems,” \emph{IEEE Internet Things J.}, vol. 7, no. 1, pp. 16–32, Jan. 2020.

       \bibitem{ref2}
       W. Hao et al., “Robust Design for Intelligent Reflecting Surface-Assisted MIMO-OFDMA Terahertz IoT Networks," \emph{IEEE Internet Things J.}, vol. 8, no. 16, pp. 13052-13064, Mar. 2021. 

		\bibitem{ref3}
		M. Hua, Q. Wu ``Throughput Maximization for IRS-Aided MIMO FD-WPCN with Non-Linear EH Model,’’ \emph{IEEE J. Sel. Topics Signal Process.}, vol. 16, no. 5, pp. 918-932, Aug. 2022.
		
		\bibitem{ref4}
		M. Hua, Q. Wu,``Joint Dynamic Passive Beamforming and Resource Allocation for IRS-Aided Full-Duplex WPCN’’ \emph{IEEE Trans. Wireless Commun.} vol. 21, no. 7, pp. 4829-4843, Jul. 2022.

	    \bibitem{ref5}
		G. Chen, Q. Wu, W. Chen, D. W. K. Ng, and L. Hanzo, “IRS-aided wireless powered MEC systems: TDMA or NOMA for computation offloading?,” \emph{IEEE Trans. Wireless Commun.}, vol. 22, no. 2, pp. 1201-1218, Feb. 2023.
		
	    \bibitem{ref6}
		H. Li, K. Xiong, R. Dong, P. Fan, and K. Ben Letaief, “Joint active and passive beamforming in IRS-enhanced wireless powered MEC networks,” \emph{IEEE Wireless Commun. Lett.}, vol. 11, no. 11, pp. 2285-2289, Nov. 2022, doi: 10.1109/LWC.2022.3199693.
		
	    \bibitem{ref7}
		S. Mao \emph{et al}., “Computation rate maximization for intelligent reflecting surface enhanced wireless powered mobile edge computing networks,” \emph{IEEE Trans. Veh. Technol.}, vol. 70, no. 10, pp. 10820 - 10831, Oct. 2021.
		
		\bibitem{ref8}
		Q. Wu, J. Xu, and R. Schober, “IRS-assisted wireless powered NOMA: Do we really need different phase shifts in DL and UL?,” \emph{IEEE Wireless Commun. Lett.}, vol. 10, no. 7, pp. 1493 - 1497, Jul. 2021.
		
		\bibitem{ref9}
		W. Hao, J. Li, G. Sun, M. Zeng, and O. A. Dobre, "Securing reconfigurable intelligent surface-aided cell-free networks," \emph{IEEE Trans. Inf. Forensics Secur.}, vol. 17, pp. 3720-3733, Oct. 2022, doi: 10.1109/TIFS.2022.3212204.
		
	    \bibitem{ref10}
		S. Abeywickrama, R. Zhang, Q. Wu, and C. Yuen, “Intelligent reflecting surface: Practical phase shift model and beamforming optimization,”  \emph{IEEE Trans. Commun.}, vol. 68, no. 9, pp. 5849-5863, Sep. 2020, doi: 10.1109/TCOMM.2020.3001125.
		
	    \bibitem{ref11}
		T. Bai et al., “Latency minimization for intelligent reflecting surface aided mobile edge computing ”, \emph{IEEE J. Sel. Areas Commun.}, vol. 38, no. 11, pp. 2666-2682, Nov. 2020.
		
		\bibitem{ref12}
		S. Mao, X. Chu, Q. Wu, L. Liu, and J. Feng, “Intelligent reflecting surface enhanced D2D cooperative computing,” \emph{IEEE Wireless Commun. Lett.}, vol. 10, no. 7, pp. 1419–1423, Jul. 2021.
		
		\bibitem{ref13}
		Z. Chu \emph{et al.}, “Intelligent reflecting surface assisted mobile edge computing for internet of things”, vol. 10, no. 3, pp. 619 - 623, Mar. 2021. 
		
		\bibitem{ref14}
		C. Sun, W. Ni, Z. Bu, and X. Wang, “Energy minimization for intelligent reflecting surface-assisted mobile edge computing,” \emph{IEEE Trans. Wireless Commun.}, vol. 21, no. 8, pp. 6329-6344, Aug. 2022.
		
		\bibitem{ref15}
		S. Huang, S. Wang, R. Wang, M. Wen, and K. Huang, “Reconfigurable intelligent surface assisted mobile edge computing with heterogeneous learning tasks,” \emph{IEEE Transactions on Cognitive Communications and Networking}, vol. 7, no. 2, pp. 369-382, Jun. 2021.
		
		\bibitem{ref16}
		 Y. Liu \emph{et al.}, “Intelligent reflecting surface meets mobile edge computing: Enhancing wireless communications for computation offloading,” arXiv:2001.07449.
		
		\bibitem{ref17}
		P. Chen, B. Lyu, and Z. Yang, “Intelligent reflecting surface enhanced wireless powered mobile edge computing,”  \emph{2021 IEEE/CIC international conference on communications in china (ICCC)}, Xiamen, China, 2021, pp. 1101-1106, doi: 10.1109/ICCC52777.2021.9580207.
		
	    \bibitem{ref18}
		Z. Chu \emph{et al.}, “Utility maximization for IRS assisted wireless powered mobile edge computing and caching (WP-MECC) networks,” \emph{IEEE Trans. Commun.}, vol. 71, no. 1, pp. 457-472, Jan. 2023, doi: 10.1109/TCOMM.2022.3222353.
		
	   \bibitem{ref19}
	   Y. Li, F. Wang, X. Zhang, and S. Guo, “IRS-based MEC for delay-constrained QoS over RF-powered 6G mobile wireless networks,” \emph{IEEE Trans. Veh. Technol.}, pp. 1-6, Jan. 2023, doi: 10.1109/TVT.2023.3234724.
		
		\bibitem{ref20}
		G. Li, M. Zeng, D. Mishra, L. Hao, Z. Ma, and O. A. Dobre, “Latency minimization for IRS-aided NOMA MEC systems with WPT-enabled IoT devices,” \emph{IEEE Internet Things J.}, vol. 10, no. 14, pp. 12156-12168, Jan. 2023, doi: 10.1109/JIOT.2023.3240395.
		
		\bibitem{ref21}
		P. Chen, B. Lyu, Y. Liu, H. Guo, and Z. Yang, “Multi-IRS assisted wireless-powered mobile edge computing for internet of things,” \emph{IEEE trans. Green Commun. Netw.}, vol. 7, no. 1, pp. 130-144, Mar. 2023, doi: 10.1109/TGCN.2022.3205030.
	
		\bibitem{ref22}
		X. Pang, N. Zhao, J. Tang, C. Wu, D. Niyato, and K. -K. Wong, ``IRS-Assisted Secure UAV Transmission via Joint Trajectory and Beamforming Design," \textit{ IEEE Trans. Commun. }, vol. 70, no. 2, pp. 1140-1152, Feb. 2022, doi: 10.1109/TCOMM.2021.3136563.
		
		\bibitem{ref23}
		X. Pang, W. Mei, N. Zhao, and R. Zhang, ``Intelligent Reflecting Surface Assisted Interference Mitigation for Cellular-Connected UAV," \textit{IEEE Wireless Commun. Lett.}, vol. 11, no. 8, pp. 1708-1712, Aug. 2022, doi: 10.1109/LWC.2022.3175920.
		
		\bibitem{ref24}
		X. Pang, W. Mei, N. Zhao, and R. Zhang, ``Cellular Sensing via Cooperative Intelligent Reflecting Surfaces," \textit{ IEEE Trans. Veh. Technol.}, pp. 1-6, Jun. 2023, doi: 10.1109/TVT.2023.3283270.
		
		\bibitem{ref25}
		C. You, K. Huang, and H. Chae, “Energy efficient mobile cloud computing powered by wireless energy transfer,” \emph{IEEE J. Sel. Areas Commun.}, vol. 34, no. 5, pp. 1757–1771, May. 2016.
		
		\bibitem{ref26}
		F. Wang, J. Xu, X. Wang, and S. Cui, “Joint offloading and computing optimization in wireless powered mobile-edge computing systems,” \emph{IEEE Trans. Wireless Commun.}, vol. 17, no. 3, pp. 1784–1797, Mar. 2018.
		
		\bibitem{ref27}
		S. Mao, S. Leng, S. Maharjan, and Y. Zhang, “Energy efficiency and delay tradeoff for wireless powered mobile-edge computing systems with multi-access schemes,” \emph{IEEE Trans. Wireless Commun.}, vol. 19, no. 3, pp. 1855–1867, Mar~2020.
		
		\bibitem{ref28}
		F. Wang, J. Xu, X. Wang, and S. Cui, “Joint offloading and computing optimization in wireless powered mobile-edge computing systems,” \emph{IEEE Trans. Wireless Commun.}, vol. 17, no. 3, pp. 1784–1797, Mar.~2018.
		
	    \bibitem{ref29}
		T. Bai, C. Pan, H. Ren, Y. Deng, M. Elkashlan, and A. Nallanathan, “Resource allocation for intelligent reflecting surface aided wireless powered mobile edge computing in OFDM systems,” \emph{IEEE Trans. Wireless Commun.}, vol. 20, no. 8, pp. 5389-5407, Aug. 2021, doi: 10.1109/TWC.2021.3067709.	
		
		\bibitem{ref30}
		Q. Wu and R. Zhang, “Towards smart and reconfigurable environment: Intelligent reflecting surface aided wireless network,” \emph{IEEE Commun. Mag.}, vol. 58, no. 1, pp. 106–112, Jan. 2020.
		
		\bibitem{ref31}
		Y. Zeng, B. Clerckx, and R. Zhang, “Communications and signals design for wireless power transmission,” \emph{IEEE Trans. Commun.}, vol. 65, no. 5, pp. 2264–2290, May 2017.
		
		\bibitem{ref32} 
		S. Bi, C. K. Ho, and R. Zhang, “Wireless powered communication: Opportunities and challenges,” \emph{IEEE Commun. Mag.}, vol. 53, no. 4, pp. 117–125, Apr. 2015.
		 
		\bibitem{ref33}
		J. Xu, L. Liu, and R. Zhang, “Multiuser MISO beamforming for simultaneous wireless information and power transfer,” \emph{IEEE Trans. Signal Process.}, vol. 62, no. 18, pp. 4798–4810, Sep. 2014.
		
		\bibitem{ref34}
		F. Wang, T. Peng, Y. Huang, and X. Wang, “Robust transceiver optimization for power-splitting based downlink MISO SWIPT systems,” \emph{IEEE Signal Process. Lett.}, vol. 22, no. 9, pp. 1492–1496, Sep. 2015.
		
		\bibitem{ref35}
		F. Wang, C. Xu, Y. Huang, X. Wang, and X.-Q. Gao, “REEL-BF design: Achieving the SDP bound for downlink beamforming with arbitrary shaping constraints,” \emph{IEEE Trans. Signal Process.}, vol. 65, no. 10, pp. 2672–2685, May 2017.
		
		\bibitem{ref36}
		J. Xu and R. Zhang, “Energy beamforming with one-bit feedback,” \emph{IEEE Trans. Signal Process.}, vol. 62, no. 20, pp. 5370–5381, Oct. 2014.
		
		\bibitem{ref37}
		Y. Wang, M. Sheng, X. Wang, L. Wang, and J. Li, “Mobile edge computing: Partial computation offloading using dynamic voltage scaling,” \emph{IEEE Trans. Commun.}, vol. 64, no. 10, pp. 4268–4282, Oct. 2016.
		
		\bibitem{ref38}
		L. Ji and S. Guo, “Energy-efficient cooperative resource allocation in wireless powered mobile edge computing,” \emph{IEEE Internet Things J.}, vol. 6, no. 3, pp. 4744–4754, Jun. 2019.
		
		\bibitem{ref39}
		L. Huang, S. Bi, and Y.-J. A. Zhang, “Deep reinforcement learning for online computation offloading in wireless powered mobile-edge computing networks,” \emph{IEEE Trans. Mobile Comput.}, vol. 19, no. 11, pp. 2581–2593, Nov. 2020.
		
		\bibitem{ref40}
		S. Mao et al., ``Computation Rate Maximization for Intelligent Reflecting Surface Enhanced Wireless Powered Mobile Edge Computing Networks," \textit{ IEEE Trans. Veh. Technol.}, vol. 70, no. 10, pp. 10820-10831, Oct. 2021, doi: 10.1109/TVT.2021.3105270.
		
		\bibitem{ref41}
		X. Hu, K. K. Wong, and Y. Zhang, ``Wireless-powered edge computing with cooperative UAV: Task, time scheduling and trajectory design,” \textit{IEEE Trans. Wireless Commun.}, vol. 19, no. 12, pp. 8083–8098, Dec. 2020.
		
		\bibitem{ref42}
		 X. Hu, K. Wong, K. Yang, and Z. Zheng, ``UAV-assisted relaying and edge computing: Scheduling and trajectory optimization,” \textit{ IEEE Trans. Wireless Commun.}, vol. 18, no. 10, pp. 4738–4752, Oct. 2019.	
		
		\bibitem{ref43}
		B. Lyu, P. Ramezani, D. T. Hoang, S. Gong, Z. Yang, and A. Jamalipour, “Optimized energy and information relaying in self-sustainable IRS-empowered WPCN,” \emph{IEEE Trans. Commun.}, vol. 69, no. 1, pp. 619–633, Jan. 2021.
		
		\bibitem{ref44}
		Y. Zou, S. Gong, J. Xu, W. Cheng, D. T. Hoang, and D. Niyato, “Wireless powered intelligent reflecting surfaces for enhancing wireless communications,” \emph{IEEE Trans. Veh. Technol.}, vol. 69, no. 10, pp. 12 369–12 373, Oct. 2020.
		
	    \bibitem{ref45}
		N. Huang, T. Wang, Y. Wu, S. Bi, L. Qian, and B. Lin “Delay minimization for intelligent reflecting surface assisted federated learning,” \emph{China Commun.}, vol. 19, no. 4, pp. 216–229, Apr. 2022.
		
		\bibitem{ref46}
		K. Shen and W. Yu, “Fractional programming for communication systems—Part I: Power control and beamforming,” \emph{IEEE Trans. Signal Process.}, vol. 66, no. 10, pp. 2616–2630, May 2018.
		
		\bibitem{ref47}
		Z. Zhang and L. Dai, “A joint precoding framework for wideband reconfigurable intelligent surface-aided cell-free network," \emph{IEEE Trans. Signal Process.}, vol. 69, pp. 4085-4101, Jun. 2021, doi: 10.1109/TSP.2021.3088755.
		
		\bibitem{ref48}
		Q. Wu and R. Zhang, “Intelligent reflecting surface enhanced wireless network via joint active and passive beamforming,” \emph{IEEE Trans. Wireless Commun.}, vol. 18, no. 11, pp. 5394–5409, Nov. 2019.        
	\end{thebibliography}
\end{document}